\documentclass[12pt]{article}
\usepackage{amsmath,amsfonts,amssymb}

\usepackage{mathtools}
\usepackage{hyperref}
\usepackage{placeins}
\usepackage[bbgreekl]{mathbbol}
\usepackage{graphicx}
\usepackage{color}
\usepackage{epsfig}
\usepackage{psfrag}	
\usepackage{tikz}
\usetikzlibrary{snakes}
\usepackage{slashed}
\usepackage{ulem}
\usepackage[font=small,labelfont=bf,width=1\textwidth]{caption}
\usepackage{comment}
\usepackage[all]{xy}
\usepackage{multirow}
\usepackage{float}
\usepackage{tensor}
\usepackage {multirow}
\usepackage {booktabs}
\usepackage {fancyhdr}
\usepackage[T1]{fontenc}

\usepackage{color}

\newcommand \be{\begin{eqnarray}}
\newcommand \ee{\end{eqnarray}}
\newcommand{\ft}[2]{{\textstyle\frac{#1}{#2}}}
\numberwithin{equation}{section}

\DeclarePairedDelimiter\floor{\lfloor}{\rfloor}

\DeclareMathOperator{\Tr}{Tr}

\def\osp{{\mathfrak{osp}}}
\def\su{{\mathfrak{su}}}
\def\u{{\mathfrak{u}}}

\newcommand{\bea}{\begin{eqnarray}}
\newcommand{\eea}{\end{eqnarray}}
\newcommand{\beq}{\begin{equation}}
\newcommand{\eeq}{\end{equation}}
\newcommand{\bal}{\begin{equation}\begin{aligned}}
\newcommand{\eal}{\end{aligned} \end{equation}}

\newcommand{\vev}[1]{{\left< {#1} \right>}}

\newcommand{\bra}[1]{{\left< {#1} \right|}}
\newcommand{\ket}[1]{{\left| {#1} \right>}}
\newcommand{\abs}[1]{{\left| {#1} \right|}}

\newcommand{\address}[1]{\vbox{\center\em#1}}
\renewcommand{\title}[1]{\vbox{\center\huge{#1}}\vspace{5mm}}

\newcommand{\cD}{{\mathcal D}}

\newcommand{\cG}{{\mathcal G}}

\newcommand{\cL}{{\mathcal L}}

\newcommand{\cN}{{\mathcal N}}

\newcommand{\cW}{{\mathcal W}}

\newcommand{\cZ}{{\mathcal Z}}

\newcommand{\ec}{\,,}

\begin{document}

\begin{titlepage}
\begin{center}

\vspace*{20mm}

\title{Bootstrap of the defect 1/2 BPS Wilson lines \\
in ${\cal N}=4$ Chern-Simons-matter theories}

\vspace{5mm}

\renewcommand{\thefootnote}{$\alph{footnote}$}

Riccardo Giordana Pozzi and Diego Trancanelli%

\vskip 3mm

\address{
Dipartimento di Scienze Fisiche, Informatiche e Matematiche, \\
Universit\`a di Modena e Reggio Emilia, via Campi 213/A, 41125 Modena, Italy \\ \& \\
INFN Sezione di Bologna, via Irnerio 46, 40126 Bologna, Italy}

\vskip 5mm

\tt{riccardo.pozzi@unimore.it, diego.trancanelli@unimore.it}

\renewcommand{\thefootnote}{\arabic{footnote}}
\setcounter{footnote}{0}

\end{center}

\vspace{8mm}
\abstract{
\normalsize{
\noindent
We compute correlation functions of local operator insertions on the 1/2 BPS Wilson lines of ${\cal N}=4$ Chern-Simons-matter theories in 3 dimensions. We study the algebra preserved by the  defect CFT supported on the line, identify the superdisplacement multiplet and discuss some of its weak-coupling realizations. By employing a superspace description, we present the 4-point functions of the superdisplacement and show how they are determined by functions of cross-ratios. Within an analytic bootstrap approach, we derive these functions at leading and next-to-leading order at  strong coupling, obtaining a result in agreement with appropriate orbifolds of the ABJM case considered in {\tt arXiv:2004.07849}.
}}
\vfill

\end{titlepage}
\tableofcontents

\section{Introduction and summary} 
\label{sec:intro}

Over the years, BPS Wilson loops have provided a rich laboratory for the investigation of superconformal theories in various dimensions. Particularly important is their role in holography, where they can be computed at strong coupling through the mapping to minimal surfaces (and other dual objects) \cite{Maldacena:1998im}, and in supersymmetric localization \cite{Pestun:2016zxk}, which in some instances allows for exact results. 

More recently, starting from \cite{Cooke:2017qgm, Giombi:2017cqn,Giombi:2018qox,Giombi:2018hsx}, there has been considerable interest in the theories defined on the contours of BPS Wilson loops, which are then regarded as 1-dimensional defects immersed in a bulk superconformal theory. The resulting defect CFTs (dCFTs) can be studied with a variety of approaches, from perturbation theory and integrability to  bootstrap techniques and holography, when a dual is available. 

The literature for 4-dimensional bulk theories is already quite extensive, besides the references cited above see also {\it e.g.} \cite{Liendo:2018ukf,Gimenez-Grau:2019hez,Correa:2019rdk,Beccaria:2019dws,Beccaria:2021rmj,Ferrero:2021bsb,Galvagno:2021bbj,Barrat:2021yvp,Barrat:2021tpn,Barrat:2022eim,Cavaglia:2022yvv,Beccaria:2022bcr,Cuomo:2021rkm,Aharony:2022ntz,Billo:2023ncz,Billo:2024kri} and \cite{Peveri:2023qip} for a review.  
The same is not true for the 3-dimensional case which has received far less attention.\footnote{See \cite{Penati:2021tfj} for a review of superconformal line defects in 3 dimensions and \cite{Castiglioni:2022yes,Castiglioni:2023uus,Castiglioni:2023tci} for recent work on RG flows in this context.} In general, Wilson loops in 3-dimensional Chern-Simons-matter theories are more complicated than their 4-dimensional counterparts, due to the natural coupling with all the fields of the theory, both bosonic and fermionic, which are organized in superconnections closely reflecting the quiver nature of these theories \cite{Drukker:2009hy}. The dCFT living on the 1/2 BPS Wilson line in ABJ(M) theory \cite{Aharony:2008ug,Aharony:2008gk} has been studied in \cite{Bianchi:2020hsz}, employing both an analytic bootstrap approach and Witten diagrams in holography. 

Inspired by the analysis in \cite{Bianchi:2020hsz}, here we consider the dCFT on the 1/2 BPS Wilson lines \cite{Ouyang:2015qma,Cooke:2015ila,Ouyang:2015iza,Ouyang:2015bmy,Mauri_2017,Mauri:2018fsf} of 3-dimensional $\mathcal{N} = 4$ Chern-Simons-matter theories (SCSM) \cite{Gaiotto:2008sd,Imamura:2008dt,Hosomichi:2008jd,Hama:2011ea}, see chapter 9 of \cite{Drukker:2019bev} for a review. These theories have superconformal algebra $\mathfrak{osp}(4|4)$, which gets broken down to ${\mathfrak{u}(1)_{j_0}}\rtimes\mathfrak{psu}(1,1|2)\rtimes\mathfrak{u}(1)_{\text{aut}}$ by the insertion of a 1/2 BPS line operator  \cite{Agmon:2020pde}. Deformations of the contour can be associated with the insertion of local operators \cite{Semenoff:2004qr,Cooke:2017qgm}, which correspond, in turn, to the broken generators of $\mathfrak{osp}(4|4)$ via a Ward identity \cite{Bianchi:2020hsz}. These local operators get arranged in a supermultiplet, the so-called superdisplacement, with a superprimary $\mathbb{R}$ of dimension 1 and its descendents $\mathbb{\Lambda}^a$ and $\mathbb{D}$ of dimensions $3/2$ and 2, respectively.

There are several SCSM theories with an $\osp(4\vert 4)$ algebra and a 1/2 BPS Wilson line \cite{Drukker:2020dvr}. They differ in the specific structure of the quiver one considers and the matter fields connecting  the nodes of the quiver. This implies that one can find different weak-coupling realizations of the superdisplacement multiplet, depending on the bulk theory considered. At the level of the correlation functions, the explicit reference to the bulk theory is encoded in certain physical quantities such as the normalization $C_{\Phi}$ of the 2-point functions, which is also related to the Bremsstrahlung function of the theory. 

These quantities are functions of the 't Hooft couplings and are explicitly related to the specific bulk theory one considers. Moreover, there exists a classical degeneracy of 1/2 BPS lines, first identified in \cite{Cooke:2015ila}, which is lifted at the quantum level \cite{Bianchi:2016vvm}, with the correct quantum operator to match the corresponding localization result being, in fact, the average of the two 1/2 BPS operators found in \cite{Cooke:2015ila}. However, the functional form of the correlation functions is fixed by the 1-dimensional symmetry \cite{Cooke:2017qgm}, which is the same for all cases, so that our conclusions will apply to all theories with ${\mathfrak{u}(1)_{j_0}}\rtimes\mathfrak{psu}(1,1|2)\rtimes\mathfrak{u}(1)_{\text{aut}}$ (we shall see that the really important bit of this algebra is $\mathfrak{psu}(1,1|2)$).  For example, the 4-point functions of the superprimary $\mathbb{R}(t)$ and its conjugate $\bar{\mathbb{R}}(t)$ are given by
\begin{align}
    \langle \mathbb{R}(t_{1})\bar{\mathbb{R}}(t_{2})\mathbb{R}(t_{3})\bar{\mathbb{R}}(t_{4}) \rangle = \frac{C_{\Phi}^{2}}{t_{12}^{2}t_{34}^{2}}f(z)
   , \qquad 
   \langle \mathbb{R}(t_{1})\bar{\mathbb{R}}(t_{2})\bar{\mathbb{R}}(t_{3})\mathbb{R}(t_{4}) \rangle
= \frac{C_{\Phi}^{2}}{t_{12}^{2}t_{34}^{2}}h(\chi),
\end{align}
with $f(z)$ and $h(\chi)$ being two functions of the cross-ratios $z$ and $\chi$ of the four coordinates of the insertions.

The main scope of this analysis is then to obtain the functions $f(z)$ and $h(\chi)$, which can be done using an analytic bootstrap approach. First, one studies the OPE for the operators in the superdisplacement multiplet, obtaining certain selection rules and identifying the protected spectrum of exchanged operators. This ultimately introduces non-trivial constraints in the bootstrap procedure. Then, one expands the functions $f(z)$ and $h(\chi)$ in conformal partial waves (CPWs), with conformal blocks that turn out to be given by $(-z)^{\Delta}\, {}_{2}F_{1}(\Delta,\Delta,2\Delta+2;z)$ for the chiral-chiral channel and $\chi^{\Delta}\, {}_{2}F_{1}(\Delta,\Delta,2\Delta;\chi)$ for the chiral-antichiral one.

The coefficients of the expansions can be found by imposing consistency conditions like crossing symmetry and mild behavior for the anomalous dimensions. One obtains in this way the expression for $f(z)$ and $h(\chi)$, as well as the anomalous dimensions, in an expansion around strong coupling with a certain expansion parameter $\epsilon$. The symmetries of the problem do not allow, however, to completely fix all coefficients and one is left with a free parameter, which we call $\xi$. This is not unexpected and in fact it is something that also takes place in other non-maximally symmetric cases, like the 1/2 BPS Wilson line in 4-dimensional $\mathcal{N}=2$~\cite{Gimenez-Grau:2019hez} super Yang-Mills theory.

One can still fix $\xi$ indirectly, by requiring consistency with the corresponding correlation function of the operator in the ABJM superdisplacement multiplet \cite{Bianchi:2020hsz} which has the same charges as our displacement operator $\mathbb{D}$. This comparison can be motivated by the fact that some ${\cal N}=4$ SCSM theories can be obtained by appropriate quotients of the ABJM quiver and it fixes the free parameter to $\xi=-\frac{3}{2}$. The final result simplifies significantly, leading to 
 \begin{equation}
     {f}(z)=1+z^2 -{4\epsilon}\Bigl[1 -\frac{z}{2} +z^2 + \frac{(1-z)}{z}\Bigl(z^3 \log(-z) +\left(1-z^3\right)\log (1-z)\Bigr)\Bigr]+ O(\epsilon^2),
     \end{equation}
and a similar expression for $h(\chi)$. Through this analysis one can also fix the anomalous dimensions obtaining 
\begin{equation}
    \Delta_n = 2+ n - \epsilon ( n^2+5n+4)+ O(\epsilon^2),
\end{equation}
for the chiral-chiral channel and
\begin{equation}
    \Delta_n = 2+ n - \epsilon ( n^2+3n)+ O(\epsilon^2), \qquad\text{$n$ even,}
\end{equation}
for the chiral-antichiral case. From a similar discussion, one can also identify the leading and next-to-leading order correction of the 4-point correlation function of the superprimary for the 1/2 BPS Wilson line in 3-dimensional $\cN=2$ theories.


The paper is organized as follows. In section \ref{sec: SUperconformal line defect} we introduce the dCFT associated with the 1/2 BPS Wilson lines of ${\cal N}=4$ Chern-Simons-matter, the preserved superconformal algebra and the corresponding superdisplacement multiplet. In section \ref{sec:correlation functions} we write down the 2- and 4-point correlation functions of this supermultiplet in terms of two functions of the cross-ratios, we obtain the selection rules for the operators in the OPE and derive the conformal blocks. Finally, in section \ref{sec:bootstrap} we perform an analytic bootstrap to get the two functions of the cross-ratios and the anomalous dimensions at strong coupling. We relegate some details about the superalgebras, their representations, and the orthogonality conditions for the coefficients of the block expansions to a series of appendices.
\section{Superconformal line defect}
\label{sec: SUperconformal line defect}

The main character of this paper is the 1-dimensional superconformal theory living on 1/2~BPS Wilson lines of 3-dimensional $\cN=4$ super Chern-Simons-matter (SCSM) theories. We provide some details about various bulk theories in the following, but for the moment it suffices to recall that these line operators are given by the path-ordered holonomy of a superconnection $\mathcal{L}(t)$\footnote{The definition of these Wilson loops in terms of superconnections, rather than ordinary bosonic connections, is characteristic of 3-dimensional quiver theories, as initially discovered for ABJM \cite{Aharony:2008ug,Aharony:2008gk} in \cite{Drukker:2009hy}. See chapter 2 of \cite{Drukker:2019bev} for a review.}
\begin{equation}
\label{WL}
{\cW}=\text{Tr}\Bigl[\mathcal{P}\exp\Bigl(-i\int_{-\infty}^{\infty}\mathcal{L}(t)\;dt\Bigr)\Bigr],
\end{equation}
with contour extending along the Euclidean time direction $t$. As shown in \cite{Ouyang:2015qma,Cooke:2015ila,Ouyang:2015iza,Ouyang:2015bmy,Mauri_2017,Mauri:2018fsf,Drukker:2020dvr,Drukker:2022ywj}, there exist various realizations of such non-local observables. Depending on the specific bulk theory, one can find an expression for $\cL(t)$ in terms of the bulk fields, for which half of the supercharges are preserved. Moreover, a 1-dimensional conformal algebra is also preserved by the operators, so that in total one has a $\u(1)_{j_0}\rtimes\mathfrak{psu}(1,1\vert 2)\rtimes \u(1)_{\text{aut}}$ defect CFT \cite{Agmon:2020pde}, as we shall see momentarily. 

We are interested in considering local insertions $\mathcal{O}_i(t_i)$ on the line and in studying their defect correlation functions
\begin{equation}
\label{defectCFTcorrfct}
\langle\mathcal{O}_1(t_1)\mathcal{O}_2(t_2)\ldots \mathcal{O}_n(t_n)\rangle_\mathcal{W}
\equiv  
\frac{\langle \Tr \mathcal{P}[ \mathcal{W}_{-\infty,t_1}\mathcal{O}_1(t_1) \mathcal{W}_{t_1,t_2}\mathcal{O}_2(t_2) \ldots \mathcal{O}_n(t_n)\mathcal{W}_{t_n,\infty}]\rangle}{\langle \cW \rangle},
\end{equation}
where $\mathcal{W}_{t_n,t_{n+1}}$ is the untraced Wilson link  $\mathcal{W}_{t_n,t_{n+1}}=\exp\Bigl(-i\int_{t_n}^{t_{n+1}}\mathcal{L}(t)dt \Bigr)$ connecting two consecutive insertions. We take the Wilson lines in the fundamental representation, so that the local operators are in the adjoint representation of the (super)gauge group in which ${\cal L}(t)$ transforms \cite{Drukker:2009hy}. The trace in \eqref{defectCFTcorrfct} guarantees a gauge invariant operator. In the following we shall drop the subscript ${\cal W}$ from the definition of the correlation functions in \eqref{defectCFTcorrfct}, to simplify the notation, and simply write $\langle\mathcal{O}_1(t_1)\mathcal{O}_2(t_2)\ldots \mathcal{O}_n(t_n)\rangle$.


\subsection{The superalgebra preserved by the defect}
\label{subsec:the preserved algebra}

The superconformal group of 3-dimensional $\mathcal{N}=4$ theories is $OSp(4\vert 4)$, see appendix \ref{app:algebra} for details. This has a $SO(1,4)\times SO(4)_{R}$ bosonic subgroup, where $SO(1,4)$ is the 3-dimensional conformal group and $SO(4)_{R}\simeq SU(2)_A \times SU(2)_B$ is the R-symmetry group. 

Turning on a straight line defect partially breaks this supergroup. In $\mathbb{R}^3$ a codimension-$2$ defect is expected to preserve $SU(1,1)\times U(1)_{\text{rot}}$, {\it i.e.} the conformal group along the line and the rotations in the plane orthogonal to the line. The insertion of the line also breaks at least half of the supercharges of the bulk theory and, consequently, this leads to only $SU(2)_A \times U(1)_B$ surviving out of the original R-symmetry. By studying carefully the breaking pattern~\cite{Agmon:2020pde}, one finds the preserved superalgebra to be $\mathfrak{u}(1)_{j_0}\rtimes \mathfrak{psu}(1,1\vert 2)\rtimes \u(1)_{\text{aut}}$, where the Abelian contributions are given by linear combinations of the $\mathfrak{u}(1)_{\text{rot}}$ and $\mathfrak{u}(1)_{B}$ generators. In particular, $\u(1)_{\text{aut}}$ represent an outer-automorphism, while $\u(1)_{j_0}$ is  a central ideal, therefore, one can simply consider $\mathfrak{psu}(1,1\vert 2)\simeq \su(1,1\vert 2)/\u(1)_{j_0}$, which we now describe in detail.

The conformal generators on the line are the translations $P$, the dilatations $K$ and the special conformal transformations $K$, which obey
\begin{align}
 [P,K]=-2 D, \qquad [D,P]=&P, \qquad [D,K]=-K.
\end{align}
The $\mathfrak{su}(2)_A$ R-symmetry generators $R_a{}^b$ (with $a,b=1,2$) obey instead
\begin{align}
 [R_a{}^b, R_c{}^d] = \delta_{c}^{b} R_a{}^d &- \delta_{a}^{d} R_c{}^b.
\end{align}
The preserved supercharges are taken to be $Q_a, \bar Q^a$ and $S_a,\bar S^a$ (see appendix \ref{app:algebra} for details) with anticommutators
\begin{align}
 \{Q_a , \bar{Q}^b\} &= 2\delta_a^b P ,\qquad  \{S_a, \bar{S}^b\} = 2\delta_a^b K,\cr
 \{Q_{a},\bar{S}^{b}\}&= 2\delta^{b}_{a}\left( D +J_{0} \right) - 2R_{a}{}^{b},\qquad
 \{\bar{Q}^{a},S_{b}\}= 2\delta^{a}_{b}\left(D - J_{0} \right)+ 2R_{b}{}^{a},
\end{align}
where
\bea
 J_{0}= i M_{12}-\bar{R}^{\dot{1}}{}_{\dot{1}}
\eea
is the $\u(1)_{j_0}$ generator given by the combination of the rotations $M_{12}$ in the orthogonal plane to the line and the $\mathfrak{u}(1)_B$ R-symmetry generator $\bar{R}^{\dot{1}}{}_{\dot{1}}$. The remaining commutation relations between bosonic and fermionic generators are reported in the appendix in \eqref{mixedFB1} and \eqref{mixedFB2}.

As it will be needed in the following, one can also work out a differential representation for the  algebra \cite{Bianchi:2020hsz, Liendo:2015cgi}. By taking superspace coordinates $(t, \theta^a, \bar\theta_a)$, where $t$ is the coordinate on the Wilson line and $\theta^a$ and $ \bar\theta_a$ are anticommuting, one obtains for $ \mathfrak{psu}(1,1\vert 2)$
\begin{align}
    & P = -\partial_t, \qquad
    D = -t \partial_t-\frac{1}{2}\theta^{a}\partial_a-\frac{1}{2}\Bar{\theta}_a\Bar{\partial}^a-\Delta\;,\cr
    & K = -t^2 \partial_t -(t+\theta\Bar{\theta})\theta^a\partial_a-(t-\theta\Bar{\theta})\Bar{\theta}_a\Bar{\partial}^a-(\theta\Bar{\theta})^2\partial_t-2 t\Delta\;,\cr
    & Q_a = \partial_a-\Bar{\theta}_a\partial_t, \qquad 
    \Bar{Q}^a = \Bar{\partial}^a-\theta^a\partial_t\;,\cr
    & S_a = (t+\theta\Bar{\theta})\partial_a-(t-\theta\Bar{\theta})\Bar{\theta}_a\partial_t-2\Bar{\theta}_a\Bar{\theta}_b\Bar{\partial}^b-2\Delta\Bar{\theta}_a\;,\cr
    & \Bar{S}^a = (t-\theta\Bar{\theta})\Bar{\partial}^a-(t+\theta\Bar{\theta}){\theta}^a\partial_t-2{\theta}^a{\theta}^b{\partial}_b-2\Delta\Bar{\theta}_a\;,\cr
    & R_a{}^{b}= -\theta^b\partial_a+\Bar{\theta}_a\Bar{\partial}^b+\frac{1}{2}\delta_a^b(\theta^c\partial_c -\Bar{\theta}_c\Bar{\partial}^c)\;,
\end{align}
where, as usual, $\partial_a = \frac{\partial}{\partial\theta^a}$, $\bar\partial^a = \frac{\bar\partial}{\bar\partial\bar\theta_a}$, the contractions are $\theta\bar\theta = \theta^a\bar\theta_a$, and $\Delta$ is the conformal dimension on the operator on which the transformations act. Moreover, as it will be needed in section \ref{sec:correlation functions}, one can also write down the quadratic Casimir for $\mathfrak{psu}(1,1\vert 2)$
\begin{equation}
\label{Casimir}
    \mathfrak{C}=D^2-\frac{1}{2}\{K,P\}+\frac{1}{4}[\Bar{S}^a,Q_a]+\frac{1}{4}[S_a,\Bar{Q}^a]-\frac{1}{2}R_a{}^b R_b{}^a\;.
\end{equation}


\subsection{The superdisplacement multiplet}
\label{sec:The superdisplacement multiplet}

In ordinary dCFTs, one can identify the displacement operator associated with the broken symmetry generators of translations orthogonal to the line \cite{McAvity:1993ue,Correa:2012at,Billo:2016cpy}. For supersymmetric theories, the displacement operator is accompanied by all other operators associated with the broken bulk generators, which are expected to arrange in a supermultiplet, the so-called superdisplacement multiplet.\footnote{Parts of this derivation of the superdisplacement multiplet have been also obtained in \cite{NagaokaPhD}.} 

By recalling the definition of the Wilson line \eqref{WL}, one can see that by considering infinitesimal variations it is possible to relate the broken symmetry generators $\mathbf{G}$ to defect operator insertions $\mathbb{G}(t)$ on the line \cite{Cooke:2017qgm,Cooke_2019}. This procedure leads to the Ward identity~\cite{Bianchi:2018zpb,Bianchi:2020hsz} 
\begin{equation}
\label{Ward-id}
    [\mathbf{G}, \cW ]={\delta_{\mathbf{G}}} \cW \equiv \int \mathcal{W}[\mathbb{G}(t)]dt ,
\end{equation}
to be understood as inserted in a given correlation function and with $\mathbb{G}(t)= -i \delta_\mathbf{G} {\cal L}(t)$. By exploiting \eqref{Ward-id}, all the defect operators associated with the respective broken generators can be identified and their quantum numbers can be extracted by applying super-Jacobi identities. The local defect operators can moreover be represented explicitly at weak coupling  in terms of the bulk Lagrangian fields, as we will show in the next subsection.

We begin our analysis from the two insertions associated with the broken R-symmetry generators  $\bar{{\bf R}}^{\dot{2}}{}_{\dot{1}}$ and $\bar{{\bf R}}^{\dot{1}}{}_{\dot{2}}$, for which one gets\footnote{From now on, we write the broken generators in boldface, to distinguish them more easily from the preserved ones.}
\begin{align}
   [\bar{{\bf R}}^{\dot{2}}{}_{\dot{1}}, \cW ]=\int \mathcal{W}[{\mathbb{R}}(t)]dt,\quad &\quad  [\bar{{\bf R}}^{\;\dot{1}}_{\dot{2}}, \cW ]  =\int \mathcal{W}[{\bar{\mathbb{R}}}(t)]dt.
\end{align}

Through appropriate super-Jacobi identities involving the insertions of the Wilson line, a broken and a preserved symmetry generator, one can assign quantum numbers to the operators inserted on the defect. For example, in order to read the $J_0$ charge of $\mathbb{R}$, one considers 
\begin{equation}
   [J_0,[\bar{{\bf R}}^{\dot{2}}_{\;\dot{1}}, \cW ]]+[\bar{{\bf R}}^{\dot{2}}_{\;\dot{1}},[\cW,J_0]]+[\cW,[J_0,\bar{{\bf R}}^{\dot{2}}_{\;\dot{1}}]]=0.
\end{equation}
Given that
\begin{align}
    [J_0,\cW]=0 ,\qquad [J_0,\bar{{\bf R}}^{\dot{2}}_{\;\dot{1}}] = \bar{{\bf R}}^{\dot{2}}_{\;\dot{1}},
\end{align}
one concludes that $\mathbb{R}$ has charge $1$ under the $U(1)_{j_0}$. 

Using the notation $[\Delta, j_0, j_1]$, where $\Delta$ is the conformal dimension, $j_0$ is the quantum number associated with the central ideal and $j_1$ the Dynkin label of the preserved $\mathfrak{su}(2)_A$ R-symmetry, the defect insertions are found to have charges 
\begin{align}
    {{\mathbb{R}}}:\quad [1,1,0],\qquad&\qquad {\bar{\mathbb{R}}}:\quad [1,-1,0].
\end{align}
Through a very similar discussion, all the remaining cases can be studied. For the broken supersymmetry generators $\mathcal{Q}_{+a\dot 1}=\epsilon_{ab}\mathbf{Q}^b$ and $\mathcal{Q}_{-a\dot 2}=i\bar{\mathbf{Q}}_a$, 
the corresponding insertions are
\begin{align}
   [{\bf Q}^a, \cW ] =\int \mathcal{W}[{\mathbb{\Lambda}}^a(t)]dt,\quad &\quad  [\bar{{\bf Q}}_a, \cW ] =\int \mathcal{W}[{\bar{\mathbb{\Lambda}}}_a(t)]dt,
\end{align}
with quantum numbers
\begin{align}
    {{\mathbb{\Lambda}}^a}:\quad \left[\ft{3}{2},1,1\right],\qquad&\qquad {\bar{\mathbb{\Lambda}}}_a:\quad \left[\ft{3}{2},-1,1\right].
\end{align}
Finally, the broken orthogonal translations, ${\bf P}=P_2+iP_1$ and  $\bar{{\bf P}}=P_2-iP_1$, define the displacement operators
\begin{align}
   [\mathbf{P}, \cW ] =\int \mathcal{W}[{\mathbb{D}}(t)]dt,\quad &\quad  [\bar{\mathbf{P}}, \cW ] =\int \mathcal{W}[{\bar{\mathbb{D}}}(t)]dt\;,
\end{align}
which have charges
\begin{align}
    {{\mathbb{D}}}:\quad [2,1,0],\qquad&\qquad {\bar{\mathbb{D}}}:\quad [2,-1,0]\;.
\end{align}

It is expected that the defect operators arrange in multiplets of the superconformal group and, therefore, they must be related by the action of the preserved generators $Q_a$. This leads to the relations
\begin{align}
[Q_a,\mathbb{R}]=\epsilon_{ab}\mathbb{\Lambda}^b,\qquad
    \{ Q_a,\mathbb{\Lambda}^b\}= 2\delta_a^b \mathbb{D},\qquad 
    [Q_a,\mathbb{D}]=0.
\end{align}

The action of $\bar Q^a$ on defect operators requires instead careful consideration due to the fact that $[\bar Q^a,\mathbf{G}] = 0$ for all  broken generators. Notably, the definition of the Ward identity \eqref{Ward-id} holds up to total derivatives along the defect \cite{Bianchi:2020hsz}. This freedom can be utilized to ensure that the action of the supercharges remains consistent with the superalgebra, which is achieved by imposing the super-Jacobi identities
\begin{equation}
    \{\bar Q^b,[Q_a,\mathbb{D}]\}-\{Q_a,[\mathbb{D},\bar Q^b]\}+[\mathbb{D},\{ \bar Q^b,Q_a \}]=0\;.
\end{equation}
By exploiting $\{Q_a ,\bar Q^b\}=2\delta^b_a P $ and that the differential action of $P$ is given by $-\partial_t$, one obtains
\begin{equation}
    \{Q_a,[\bar Q^b,\mathbb{D}]\}-2\delta^b_a\partial_t \mathbb{D}=\{Q_a,[\bar Q^b,\mathbb{D}]-\partial_t \mathbb{\Lambda}^b\}=0\;,
\end{equation}
leading to $ [\Bar{Q}^a,\mathbb{D}]=\partial_t \mathbb{\Lambda}^a$. For $\{\bar Q^a,\mathbb{\Lambda}^b\}$ a similar discussion applies:
\begin{align}
    [\Bar{Q}^b,\{ Q_a, \mathbb{\Lambda}^c\}]+[Q_a,\{  \mathbb{\Lambda}^c,\bar Q^b\}]+[ \mathbb{\Lambda}^c,\{\bar Q^b, Q_a\}]&=0\,,\cr
    2\delta^c_a\partial_t \mathbb{\Lambda}^b+[Q_a,\{  \mathbb{\Lambda}^c,\bar Q^b\}]-2\delta^b_a\partial_t \mathbb{\Lambda}^c&=0\,,\cr
    [Q_a,\{\bar Q^b,\mathbb{\Lambda}^c\}-2\epsilon^{bc}\partial_t \mathbb{R}]&=0\,.
\end{align}
This implies that $\{\bar Q^b,\mathbb{\Lambda}^c\}=2\epsilon^{bc}\partial_t \mathbb{R}$. Overall, one finds the following relations
\begin{equation}
 [\Bar{Q}^a,\mathbb{R}]=0, \qquad 
 \{\Bar{Q}^a, \mathbb{\Lambda}^b\}= 2\epsilon^{ab}\partial_t \mathbb{R}, \qquad
[\Bar{Q}^a,\mathbb{D}]=\partial_t \mathbb{\Lambda}^a.
\end{equation}

We see therefore that $\mathbb{R}$ ($\bar{\mathbb{R}}$) is the superconformal primary of the superconformal multiplet
\begin{equation}
\label{superdisplacementcharges}
    L\bar{A}[1]^{(0)}_{1}:\quad [1,1,0]\longrightarrow \left[\ft{3}{2},1,1\right]\longrightarrow [2,1,0],
\end{equation}
where we adopt the notation of \cite{Agmon:2020pde, Cordova:2016emh} with $[j_0]^{(j_1)}_{\Delta}$ indicating the quantum numbers of the superprimary\footnote{We mix the two equivalent notations $[j_0]^{(j_1)}_{\Delta}=[\Delta,j_0,j_1]$.} and the capital letters specifying whether the multiplet is long $L$ ($\bar L$) or short at threshold $A$ ($\bar A$) with respect to $Q_a$ ($\bar Q^a$), see also appendix~\ref{app:representations} for more details. In this specific case, $L\bar A$ means that the multiplet is long with respect to $Q_a$ and short with respect to $\bar Q^a$. Again, as expected from the central ideal of the defect algebra, all $Q$-descendants have the same $U(1)_{j_0}$ charge.

This analysis is consistent with \cite{Agmon:2020pde}, where our superdisplacement multiplet is also considered as part of their general classification of superconformal defects. 


\subsection{1/2 BPS Wilson lines in $\mathcal{N}=4$ theories}
\label{subsec:The 1/2 BPS line in}

Here we briefly review some 1/2 BPS Wilson lines in ${\cal N}=4$ theories, thus providing an explicit realization of the defects discussed so far. These operators have been initially introduced in \cite{Ouyang:2015qma,Cooke:2015ila} and then generalized and studied in, {\it e.g.} \cite{Ouyang:2015iza,Ouyang:2015bmy,Mauri_2017,Mauri:2018fsf,Drukker:2020dvr,Drukker:2022ywj}. See chapter 9 of \cite{Drukker:2019bev} for a review. 

The theories we consider are defined by circular or linear quivers with gauge groups $U(N_i)$ \cite{Gaiotto:2008sd,Hosomichi:2008jd}. The nodes, labelled by the index $i$, are connected to each others by bifundamental fields, which can be either hypers or twisted hypers. The components of the hypers (twisted hypers) are the scalars $q_{(i)a}$ ($q_{(i)}^{\dot a}$) and the fermions $\psi_{(i)}^{\dot a}$ ($\psi_{(i)a}$), all in the (anti)bifundamental representation of the gauge groups $U(N_{i})$ and $U(N_{i+1})$. We recall that $a=1,2$ is the index for $\mathfrak{su}(2)_A$, while $\dot a=\dot 1,\dot 2$ is the one for $\mathfrak{su}(2)_B$.

\subsubsection{Alternating CS levels}

Following \cite{Cooke:2015ila}, we first consider circular quivers with vector multiplets coupled to hypers and twisted hypers. In particular, we consider $\cN=4$ SCSM theories with alternating levels specified by the (partial) necklace quiver diagram
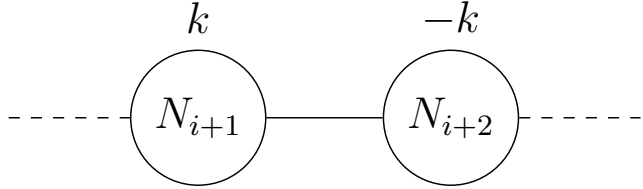
\begin{center}
\scalebox{1.4}{
\begin{tikzpicture}[every node/.style={circle,draw}]
\node(NL) at (1.8,0){$N_{i+1}$};
\node(NR) at (4.2,0){$N_{i+2}$};
\draw[dashed](0,0)--(NL);
\draw[solid](NL)--(NR);
\draw[dashed](NR)--(6,0);
\begin{scope}[nodes = {draw = none,above = 16pt}]
\node at (NL) {$k$};
\end{scope}
\begin{scope}[nodes = {draw = none,above = 12pt}]
\node at (NR) {$-k$};
\end{scope}
\end{tikzpicture}
 }
\end{center}
where the solid lines stand for hypers linking the two nodes and the dashed line for twisted hypers.
One defines the following scalar bilinears \cite{Imamura:2008dt,Cooke:2015ila}
\begin{align}
\label{mm}
&\nu_{(i)}=q_{(i)a} \bar q_{(i)}^a,
&&\tilde\nu_{(i)}=\bar q_{(i)}^a q_{(i)a}\,,
\cr
&(\mu_{(i)})_{a}{}^{b}
=q_{(i)a} \bar q_{(i)}^b
-\frac{1}{2}\delta_{a}{}^{b} \nu_{(i)},
&&
(\tilde\mu_{(i)})^{a}{}_{b}
=\bar q_{(i)}^a q_{(i)b}
-\frac{1}{2}\delta^{a}{}_{b} \tilde\nu_{(i)}\,.
\end{align}
transforming in the adjoint representation of the gauge group of their respective node.

For this case we consider the so-called $\psi_1$-loop specified by the superconnection\footnote{With respect to \cite{Cooke:2015ila} we have just added a factor of $\pi$. There is a second operator, called $\psi_2$-loop, corresponding to having $\mathfrak{su}(2)_B$ preserved. The choice of which R-symmetry subgroup is preserved and which one is broken is immaterial, being just a matter of swapping hypers and twisted hypers. There is also a sign difference in the scalar coupling, which however does not affect the result at this level.}
\begin{equation}
\label{Lpsi1}
    \mathcal{L}_{\psi_1}=
\begin{pmatrix}
A_{(i+1)1}-\frac{2\pi i}{k}(\tilde\mu_{(i)})^{\dot{1}}{}_{\dot{1}}-\frac{i\pi}{k}\nu_{(i+1)}
&{(1-i)}\sqrt{\frac{\pi}{k}}\,\psi_{(i+1)\dot{1}}^{+}\\
{(1-i)}\sqrt{\frac{\pi}{k}}\,\bar\psi_{(i+1)+}^{\;\dot{1}}
&A_{(i+2)1}-\frac{2\pi i}{k}(\mu_{(i+2)})_{\dot{1}}{}^{\dot{1}}-\frac{i\pi}{k}\tilde\nu_{(i+1)}
\end{pmatrix}
\end{equation}
defined in terms of fields in the $\{i,i+1,i+2\}$ nodes. The spinor indices are $\pm$, see appendix \ref{app:algebra} for details. The Wilson loop is then \eqref{WL}, with the superconnection above. It preserves half of the superconformal charges \cite{Cooke:2015ila}
 \begin{equation}
      Q_{+a\dot{2}}\;,\qquad S^{-}_{a\dot{2}}\;,  \qquad  Q_{- b\dot{1}}\;, \qquad S^{+}_{b\dot{1}}\;.
 \end{equation}
For example, by looking at the contributions with  $\mu_{(i+2)}$ and $\tilde \mu_{(i)}$, only the diagonal part of $\mathfrak{su}(2)_B$ is unbroken. On the other hand, the line defect has to commute with the preserved symmetry generators. For example, one can look at the relation involving $J_0$ which, consistently with the previous discussion, vanishes for the fermions
\begin{equation}
    [J_0\;,\;\psi_{(i+1)\dot{1}}^{+}]= 0,
\end{equation}
and for all other fields. 

It is worth recalling that for these Wilson lines expressed in terms of superconnection, one relaxes the requirement that the supersymmetry variation $\delta_S$ of the superconnection be zero, imposing instead the weaker requirement \cite{Drukker:2009hy,Cardinali:2012ru}
 \begin{equation}
 \label{gauge transf}
     \delta_S \cL = \cD_t \cG= \partial_t\cG + i[\cL , \cG],
 \end{equation}
 where $\cG$ is a $U(N_{i+1}\vert N_{i+2})$ supermatrix. The gauge invariance of the operator guarantees then its supersymmetry invariance as well.

We are now in the position to write down a weak-coupling realization of the superprimary operator $\mathbb{R}$ in terms of the ingredients introduced above. By computing explicitly the variation \eqref{Ward-id} of the superconnection \eqref{Lpsi1} with respect to the broken R-symmetry generators, one finds 
\begin{equation}
\label{weakcouplingrep1}
\mathbb{R} = -\begin{pmatrix}
\frac{2\pi}{k}(\Tilde{\mu}_{(i)})^{\dot{2}}_{\;\dot{1}} & 0\\[0.1cm]
(1+i)\sqrt{\frac{\pi}{k}}\Bar{\psi}_{(i+1)+}^{\;\dot{2}}\;  & \;\frac{2\pi}{k}({\mu}_{(i+2)})^{\;\dot{2}}_{\dot{1}}
\end{pmatrix},
\qquad
\bar{\mathbb{R}} =\begin{pmatrix}
\frac{2\pi}{k}({\mu}_{(i)})_{\dot{2}}{}^{\dot{1}}\;  & (1+i)\sqrt{\frac{\pi}{k}}{\psi}^{+}_{(i+1)\dot{2}}\\
 0 & \frac{2\pi}{k}(\Tilde{\mu}_{(i+2)})^{\dot{1}}{}_{\dot{2}}
\end{pmatrix}.
\end{equation}
The $Q$-descendants can be obtained in a similar way, exploiting \eqref{Ward-id} or, equivalently,  the gauge transformation \eqref{gauge transf}.


\subsubsection{Linear quivers}
The same analysis done for the case of alternating CS level can be performed analogously for all the other cases. The linear case just follows from the previous, by removing a hyper or twisted hyper from the circular case \cite{Cooke:2015ila}:
\begin{equation}
    \mathcal{L}_{\psi_1}=
\begin{pmatrix}
A_{(i+1)1}-\frac{2\pi i}{k}(\tilde\mu_{(i)})^{\dot{1}}{}_{\dot{1}}-\frac{i\pi}{k}\nu_{(i+1)}
\;\;&\;\;{(1-i)}\sqrt{\frac{\pi}{k}}\,\psi_{(i+1)\dot{1}}^{+}\\
{(1-i)}\sqrt{\frac{\pi}{k}}\,\bar\psi_{(i+1)+}^{\;\dot{1}}
&A_{(i+2)1}-\frac{i\pi}{k}\tilde\nu_{(i+1)}
\end{pmatrix}.
\end{equation}
Again, one can identify the superprimaries for this particular case, which read  
\begin{equation}
\label{weakcouplingrep2}
\mathbb{R} =-  \begin{pmatrix}
\frac{2\pi}{k}(\Tilde{\mu}_{(i)})^{\dot{2}}_{\;\dot{1}} & 0\\[0.1cm]
(1+i)\sqrt{\frac{\pi}{k}}\Bar{\psi}_{(i+1)+}^{\;\dot{2}}\;  & \;0
\end{pmatrix},
\qquad
\bar{\mathbb{R}} =\begin{pmatrix}
\frac{2\pi}{k}({\mu}_{(i)})^{\;\dot{1}}_{\dot{2}}\; &\; (1+i)\sqrt{\frac{\pi}{k}}{\psi}_{(i+1)}^{+\dot{2}}\\
 0 & 0
\end{pmatrix}.
\end{equation}

It is clear at this point that for each case of \cite{Cooke:2015ila} (or of \cite{Drukker:2020dvr,Drukker:2022ywj}) one can construct the corresponding weak coupling representation in terms of supermatrices of the bulk fields, through the prescription in \eqref{Ward-id}. It is rather interesting to note that these different 1/2 BPS Wilson lines are all mapped to the same functionally equivalent quantities in the defect CFT analysis, as dictated by the preserved superconformal symmetry. However, the explicit reference to a specific bulk theory is not be completely lost, for it is encoded in a physical normalization factor and a small expansion parameter of the correlation functions.
\section{Defect correlation functions}
\label{sec:correlation functions}

Now we organize the operators of the superdisplacement multiplet into (anti)chiral superfields, for which one can define and study correlation functions in superspace. This will set the groundwork for the next section.

\subsection{Correlation functions in superspace}

We begin by defining (anti)chiral coordinates in superspace
\begin{align}
    y= t + \theta^a \Bar{\theta}_a, \qquad 
    \bar{y} &= t - \theta^a \Bar{\theta}_a,
\end{align}
and the corresponding covariant derivatives 
\begin{equation}
    D_a = \partial_a + \Bar{\theta}_a\partial_t, \qquad  \Bar{D}^a = \Bar{\partial}^a + {\theta}^a\partial_t ,
\end{equation}
such that $D_a \bar y=0$ and $\bar{D}_a  y=0$. 
This allows to introduce the component expansion of generic (anti)chiral superfields obeying the conditions  $\bar{D}^{a}\Phi(y,\theta) = 0$ and $D_a\bar \Phi(\bar y,\bar\theta)=0$:
\begin{align}
\Phi(y,\theta) &= \phi(y) + \theta^{a}\lambda_{a}(y) +\theta^{a}\theta^{b}\epsilon_{ab} F(y),\cr
\bar{\Phi}(\bar y,\bar\theta) &= \bar{\phi}(\bar y) + \bar{\theta}_{a}\bar{\lambda}^{a}(\bar y) + \bar{\theta}_{a}\bar{\theta}_{b}\epsilon^{ab}\bar{F}(\bar y).
\end{align}
The 2-point function in superspace is then given by
\begin{equation}
\langle \Phi(y_{1},\theta_{1}) \bar{\Phi}(\bar y_{2},\bar \theta_{2})\rangle = \frac{C_{\Phi}}{\langle 1\bar{2} \rangle^{2\Delta}},
\label{2pt-superfield-N=4}
\end{equation}
where \cite{Bianchi:2020hsz}
\begin{equation}
    \langle i \bar j \rangle = y_i - \bar y_j -2\theta^a_i\bar\theta_{ja}
\end{equation}
is the chiral distance between two points in superspace and ${\Delta}$ the conformal dimension of the superfields. 

In the case of interest, the superfields associated with the superdisplacement multiplet of the 1/2 BPS Wilson line are
\begin{align}
\Phi(y,\theta)  &= \mathbb{R}(y) + \theta^{a}{\mathbb{\Lambda}}_{a}(y) +\theta^{a}\theta^{b}\epsilon_{ab} \mathbb{D}(y),\cr
\bar{\Phi}(\bar y,\bar\theta) &= \bar{\mathbb{R}}(\bar y) + \bar{\theta}_{a}\bar{\mathbb{\Lambda}}^{a}(\bar y) + \bar{\theta}_{a}\bar{\theta}_{b}\epsilon^{ab}\bar{\mathbb{D}}(\bar y),
\label{superfields}
\end{align}
and have $\Delta=1$. The 2-point functions for each defect operator are obtained by expanding \eqref{2pt-superfield-N=4} in the Grassmann variables. By comparing the product of the superfields on the left-hand side and the expansion of the chiral distance on the right-hand side, one gets\footnote{Alternatively to expanding \eqref{2pt-superfield-N=4}, one can equivalently take \eqref{2pnt-components} with generic normalizations of the various components, $C_{\mathbb{R},\mathbb{\Lambda},\mathbb{D}}$, and relate them by applying the supercharges on mixed 2-point correlation functions like $\vev{\mathbb{R}(t_1)\bar{\mathbb{\Lambda}}_a(t_2)}=0$. One finds $C_{\Phi}=C_{\mathbb{R}}=4C_{\mathbb{\Lambda}}$ and $2C_{\mathbb{D}}=3C_{\mathbb{\Lambda}}$.}
\begin{align}
\label{2pnt-components}
\langle \mathbb{R}(t_{1})\bar{\mathbb{R}}(t_{2}) \rangle = \frac{C_{\Phi}}{t_{12}^{2}},\qquad
\langle \mathbb{\Lambda}^{a}(t_{1})\bar{\mathbb{\Lambda}}_{b}(t_{2}) \rangle  =\frac{4C_{\Phi}}{t_{12}^{3}}\delta^{a}_{b},\qquad
\langle \mathbb{D}(t_{1})\bar{\mathbb{D}}(t_{2}) \rangle &= \frac{6C_{\Phi}}{t_{12}^{4}},
\end{align}
with $t_{ij}\equiv t_i-t_j$. The powers of $t_{12}$ in these expressions are of course consistent with the conformal dimensions of the components of the superdisplacement \eqref{superdisplacementcharges}. 
The normalization $C_\Phi$ depends on the coupling constant of the theory and, for the superdisplacement multiplet, it has an important physical meaning, being generally related to the Bremsstrahlung function~\cite{Correa:2012at}.\footnote{For ABJM see also, {\it e.g }, \cite{Bianchi:2014laa,Correa:2014aga,Bianchi_2017,Bianchi:2018scb,Castiglioni:2023tci} and chapters 10 and 11 of \cite{Drukker:2019bev} for a review.} 

Moving on to the 4-point functions, there are two inequivalent ordering choices:
\begin{equation}
\label{4pnt-generic-1}
\langle \Phi(y_{1},\theta_{1}) \bar{\Phi}(\bar y_{2},\bar \theta_{2}) \Phi(y_{3},\theta_{3}) \bar{\Phi}(\bar y_{4},\bar \theta_{4})\rangle = \frac{C_{\Phi}^{2}}{\langle 1\bar{2} \rangle^{2} \langle 3\bar{4} \rangle^{2}}f(\mathcal{Z}),
\end{equation}
and
\begin{equation}
\label{4pnt-generic-2}
\langle \Phi(y_{1},\theta_{1}) \bar{\Phi}(\bar y_{2},\bar \theta_{2}) \bar{\Phi}(\bar y_{3},\bar \theta_{3}) \Phi(y_{4},\theta_{4})\rangle = \frac{C_{\Phi}^{2}}{\langle 1\bar{2} \rangle^{2} \langle 4\bar{3} \rangle^{2}}h(\mathcal{X}).
\end{equation}
Here $f(\cZ)$ and $h(\mathcal{X})$ are functions of the cross-ratios of the chiral distances
\begin{equation}
\cZ = \frac{\langle 1\bar{2} \rangle \langle 3\bar{4} \rangle}{\langle 1\bar{4} \rangle \langle 3\bar{2} \rangle}, \qquad \mathcal{X} = \frac{\langle 1\bar{2} \rangle \langle 4\bar{3} \rangle}{\langle 1\bar{3} \rangle \langle 4\bar{2} \rangle}.
\end{equation}
By expanding in the fermionic coordinates, one can identify the bosonic part of these cross-ratios, which are given by
\bea
\label{crossratioszchi}
z=\frac{t_{12}t_{34}}{t_{14}t_{32}},\qquad \chi=\frac{t_{12}t_{34}}{t_{13}t_{24}}.
\eea
In the specific case of 1-dimensional theories, the operator insertions are ordered along the line: $t_{1}<t_{2}<t_{3}<t_{4}$. 

Before moving on, it is worth discussing the domains of the variables and their relations. First of all, as typical in CFTs, one can exploit the symmetries to fix a frame. In the following we will work in the conformal frame specified by $t_1\rightarrow0$, $t_3\rightarrow1$ and $t_4\rightarrow\infty$. In this setup the cross-ratios are given  by 
\begin{equation}
    z = \frac{t_2}{t_2-1}\in (-\infty,0),\qquad \chi={t_2}\in (0,1)\,,
\end{equation}
which are related by
\begin{equation} 
z = \frac{\chi}{\chi-1},
\end{equation}
with branch cut singularities at the endpoints of the intervals above. Relevant limits are given by $t_2 \rightarrow t_1$, which implies $\chi\rightarrow 0$ and $z\rightarrow 0$, and by $t_2 \rightarrow t_3$, which implies $\chi \rightarrow 1$ and $z\rightarrow -\infty$. Moreover, under the mapping $\chi\rightarrow 1-\chi$, one has  $z\rightarrow 1/z$. We will see how these limits will play a role in the following.

Similarly to the 2-point functions, one can  specialize \eqref{4pnt-generic-2} to the superdisplacement multiplet and, by expanding both sides, one can obtain the expressions in terms of the defect insertions. The most straightforward relations are the ones regarding the superprimaries, which are given by
\begin{align}
\label{superprimary 4 point correlators}
    \langle \mathbb{R}(t_{1})\bar{\mathbb{R}}(t_{2})\mathbb{R}(t_{3})\bar{\mathbb{R}}(t_{4}) \rangle = \frac{C_{\Phi}^{2}}{t_{12}^{2}t_{34}^{2}}f(z)
   , \qquad 
   \langle \mathbb{R}(t_{1})\bar{\mathbb{R}}(t_{2})\bar{\mathbb{R}}(t_{3})\mathbb{R}(t_{4}) \rangle
= \frac{C_{\Phi}^{2}}{t_{12}^{2}t_{34}^{2}}h(\chi).
\end{align}
For reference, in other to distinguish them in the following, we will refer to the first correlator as the $f$-correlator and to the second one as the $h$-correlator.

It is remarkable to notice how the preserved supersymmetry allows to determine all 4-point correlation functions of the supermultiplet insertions in terms of just two quantities, $f(z)$ and $h(\chi)$, and their derivatives. For example, the fermionic superdescendant have
\begin{align}
\langle  \mathbb{\Lambda}^{a_1}(t_{1})\bar{\mathbb{\Lambda}}_{a_2}(t_{2})  \mathbb{\Lambda}^{a_3}(t_{3})\bar{\mathbb{\Lambda}}_{a_4}(t_{4})  \rangle & = \frac{(4 C_{\Phi})^{2}}{t_{12}^{3}t_{34}^{3}}\frac{1}{4}
\Bigl[\delta^{a_1}_{a_2}\delta^{a_3}_{a_4}\Bigl( 4 f-3zf'+z^2 f''\Bigr)-\delta^{a_1}_{a_4}\delta^{a_3}_{a_2}\Bigl( z^2 f'+z^3f''\Bigr)\Bigr]
\label{fermionic 4 point function},
\end{align}
while for the displacement operators one obtains
\bea
&&
\langle \mathbb{D}(t_{1})\bar{\mathbb{D}}(t_{2})\mathbb{D}(t_{3})\bar{\mathbb{D}}(t_{4}) \rangle 
= \frac{(6 C_{\Phi})^{2}}{t_{12}^{4}t_{34}^{4}}\frac{1}{36}
\Bigl[ 36f+ 
z\left(4 z^2-2 z-32\right) f'+2 z^2 \left(7 z^2+z+7\right) f''
\cr && \hskip 7cm
+ z^3(z-1)(8 z+4) f^{(3)}+z^4(z-1)^2  f^{(4)}
\Bigr],
\label{displacement 4 point function}
\eea
with $f=f(z)$ everywhere. Similarly, all other correlators that can appear by expanding \eqref{4pnt-generic-1} or \eqref{4pnt-generic-2} are completely specified by either $f(z)$ or $h(\chi)$ and their derivatives. The computation of the correlators reduces then to determining these functions.


\subsection{Selection rules}
\label{Selection rules}

We now study the Operator Product Expansion (OPE) for the superdisplacement multiplet and obtain the selection rules that constrain the exchanged operators. Due to the similarity with ABJM, some of the considerations in \cite{Bianchi:2020hsz} may be also applied here.

Let us start with the chiral-antichiral case, namely the OPE between a chiral and an antichiral superfield. In \cite{Bianchi:2018scb} it was found that only the identity and long multiplets can appear in the chiral-antichiral OPE for the $\mathfrak{su}(1,1|1)$ case, which corresponds to ${\cal N}=2$ theories. 
It follows that in all other algebras with higher supersymmetry, one can always identify the subalgebra $\mathfrak{su}(1,1|1)\subset \mathfrak{su}(1,1|\floor{\mathcal{N}/2})$ generated by $Q_a,\Bar{Q}^a$ for fixed $a$ \cite{Bianchi:2020hsz}. This is true for the 1/2 BPS Wilson line in $\mathcal{N}=6$, but also for the one of $\mathcal{N}=4$. Hence, the OPE schematically reads
 \begin{equation}
 \label{phi-phibar-selection rule}
    L\Bar{A}[j_0]_{j_0}^{(0)} \times A\Bar{L}[-j_0]_{j_0}^{(0)}\sim \mathcal{I}+L\Bar{L}[0]_{\Delta}^{(0)},
\end{equation}
where the notation introduced at the end of section \ref{sec:The superdisplacement multiplet} and in appendix \ref{app:representations} is adopted.

Regarding the chiral-chiral case, one can study the OPE expansion by requiring consistency under the chirality condition and the charges. In particular, if one considers the OPE for $[j_0]_{j_0}^{(0)}\times [j_0]_{j_0}^{(0)}$ where $[j_0]_{j_0}^{(0)}$ is the superprimary of $LA[j_0]_{j_0}^{(0)}$, the first contribution one expects to be exchanged is again a superprimary, as it satisfies the full shortening condition, with quantum numbers $[2j_0]_{2j_0}^{(0)}$. Notice, however, that it is not the full $LA[2j_0]_{2j_0}^{(0)}$ multiplet that takes part in the OPE, but just one of its component (in this case the superprimary), with all its conformal family. Analogously, all other contributions that can be exchanged are constrained by chirality. These are built by acting a sufficient number of times with $\bar{Q}^a$ on a given superprimary, until the correct chirality condition and charge consistency are met. The result for the supermultiplets is schematically given by
\begin{equation}
    L\Bar{A}[j_0]_{j_0}^{(0)}\times L\Bar{A}[j_0]_{j_0}^{(0)}\sim L\Bar{A}[2j_0]_{2j_0}^{(0)}+L\Bar{A}[2j_0]^{(1)}_{2j_0+\frac{1}{2}}+L\Bar{L}[2j_0]^{(0)}_{\Delta}.
\end{equation}
Then, when considering the superprimary OPE, one obtains (see \eqref{chargesoftheQs} for the charges of $\bar Q^a$)
\begin{equation}
\label{phi-phi selection rule}
    [j_0]_{j_0}^{(0)}\times [j_0]_{j_0}^{(0)}\sim [2j_0]_{2j_0}^{(0)}+\bar{Q}^{1}[2j_0]^{(1)}_{2j_0+\frac{1}{2}}+\bar{Q}^{2}\bar{Q}^{1}[2j_0]^{(0)}_{\Delta},
\end{equation}
where we have explicitly written down the $\bar{Q}$-descendants given by the action of the supercharges on the corresponding superprimaries. The quantum numbers of the exchanged conformal primaries are therefore $[2j_0,2j_0,0]$, $[2j_0+1,2j_0,0]$ and $[\Delta+1, 2j_0,0]$.

In order to complete the discussion, one has also to investigate the implications of the decomposition rules at the threshold $ \Delta_\star = j_0+\frac{1}{2}j_1$ for the long superprimary. In particular, one has \cite{Agmon:2020pde}
\begin{equation}
        L\bar{L}[j_0]^{(j_1)}_{\Delta\rightarrow \Delta_\star}=  L\bar{A}[j_0]^{(j_1)}_{\Delta_\star}\oplus L\bar{A}[j_0]^{(j_1+1)}_{\Delta_\star+\frac{1}{2}},
\end{equation}
which, specialized to our case in order to meet \eqref{phi-phi selection rule}, reads
\begin{equation}
        L\bar{L}[2j_0]^{(0)}_{\Delta\rightarrow \Delta_\star}=  L\bar{A}[2j_0]^{(0)}_{2j_0}\oplus L\bar{A}[2j_0]^{(1)}_{2j_0+\frac{1}{2}}.
\end{equation}
Since this is precisely the contribution that has been already considered, one can avoid this case by requiring that the dimension of long multiplets  never hits the threshold value. Therefore, the long superprimaries are only allowed to have dimension strictly greater the the saturating one, $\Delta^{\text{long-SP}}>2j_0$. Rewritten in terms of the $\Bar{Q}$-superdescendants appearing in the expansion, this means
\begin{equation}
    \Delta^{\text{$\Bar{Q}$-des}}_{\text{long}} > 2j_0+1.
\end{equation}
Consequently, the OPE within the chiral-chiral sector allows for the exchange of two conformal primaries that are protected and have dimensions $2j_0$ and $2j_0+1$, alongside an infinite number of operators whose dimensions are unprotected and strictly larger than those of the protected operators.


\subsection{Block expansions}
\label{subsec: Superconformal blocks}

Now everything is set for identifying the superconformal blocks of the Conformal Partial Wave (CPW) expansions \cite{Dolan:2003hv} of \eqref{superprimary 4 point correlators}. These contributions will be associated with the conformal family of the operators that appear in the OPE expansion of the corresponding channel. In the following we identify all such contributions.

\subsubsection{$f$-correlator}

We start by the s-channel, in which we expand around the insertions at $(12)$ and $(34)$. In terms of the cross ratios, this corresponds to $z\rightarrow 0$. The chiral-antichiral OPE \eqref{phi-phibar-selection rule} can be exploited, thus leading to the CPW expansion of $f(z)$ given by 
\begin{equation}
\label{CPW f s-channel}
    f(z) = 1 + \sum_{\Delta>0}a_{\Delta}F_{\Delta}(z),
\end{equation}
where $\Delta>0$ follows directly from the selection rule and the blocks sum the contributions of the superconformal descendants with quantum numbers $[\Delta,0,0]$. In order to identify them, we rely on the shadow formalism \cite{Ferrara:1972uq}, see \cite{Poland:2018epd} for a review. We consider the quadratic Casimir \eqref{Casimir} acting on the operators inserted at $y_1$ and $y_2$. To this scope, we replace the generators in \eqref{Casimir} with generators acting on the two coordinates, {\it e.g.} $D\rightarrow D_{\text{s}}= D_1 +D_2$, $K\rightarrow K_{\text{s}}= K_1 +K_2$ and so on.\footnote{This is just the same as considering $\mathfrak{D}_{1,2}\propto (\mathcal{J}_{ab,1}+\mathcal{J}_{ab,2})(\mathcal{J}^{ab}_{1}+\mathcal{J}^{ab}_{2})$ where $\mathcal{J}^{ab}_{1}$ are the differential representations of the generators of the superconformal algebra and the single Casimir is $\mathfrak{C}\propto \mathcal{J}^{ab}\mathcal{J}_{ab} $} Overall, the differential operator one obtains is 
\begin{equation}
\label{Casimir diff}
    \mathfrak{D}_{1,2}=D_{\text{s}}^2-\frac{1}{2}\{K_{\text{s}},P_{\text{s}}\}+\frac{1}{4}[\Bar{S}_{\text{s}}^a,Q_{\text{s}a}]+\frac{1}{4}[S_{\text{s}a},\Bar{Q}_{\text{s}}^a]-\frac{1}{2}R_{\text{s}a}{}^b R_{\text{s}b}{}^a\;.
\end{equation}

Moreover, from \eqref{Casimir} and the quadratic Casimir eigenvalue $\mathfrak{c}_\Delta=\Delta (\Delta+1)$, one reaches the condition 
\begin{equation}
\label{casimir equation}
\left( \mathfrak{D}_{1,2} - \mathfrak{c}_\Delta \right)\langle \Phi(y_{1},\theta_{1}) \bar{\Phi}(\bar y_{2},\bar \theta_{2}) \Phi(y_{3},\theta_{3}) \bar{\Phi}(\bar y_{4},\bar \theta_{4})\rangle=0.
\end{equation}
By expanding in the Grassmann variables, one  can identify the second-order differential equation
\begin{equation}
\label{differential equation}
z (2-z)\;\partial_z F_{\Delta}(z)+(1-z) \;z^2\; \partial_z^2F_{\Delta}(z) = \Delta  (\Delta +1) F_{\Delta}(z)\;,
\end{equation}
which is solved by the linear combination\footnote{An initial derivation of these conformal blocks is contained in \cite{NagaokaPhD}.} 
\begin{equation}
   F_{\Delta}(z) = c_1 (-z)^{-\Delta -1} \, _2F_1(-\Delta -1,-\Delta -1,-2 \Delta ;z)+c_2  (-z)^{\Delta } \, _2F_1(\Delta ,\Delta ,2 \Delta +2;z), 
\end{equation}
where $c_1$ and $c_2$ are integration constants. By defining  
\begin{equation}
   G_{\Delta}(z)= (-z)^{\Delta}\, {}_{2}F_{1}(\Delta,\Delta,2\Delta+2;z),
   \label{Ghypergeometric}
\end{equation}
the superconformal blocks have the simple expression
\begin{equation}
   F_{\Delta}(z) = c_1 G_{-1-\Delta}(z)+c_2 G_{\Delta}(z)\;.
\end{equation}
The $G_{-1-\Delta}(z)$  term is associated with the so-called shadow contributions. Its appearance is related to the fact that $ G_{-1-\Delta}(z)$ and $ G_{\Delta}(z)$ have the same eigenvalues $\mathfrak{c}_\Delta = \mathfrak{c}_{-1-\Delta}$. Having dimension strictly less than the unitarity bound, they have no real physical interpretation, meaning that they do not belong to the physical spectrum of the theory. So, consistently with the OPE, one sets $c_1 =0$ and $c_2 = 1$. However, with a little foresight, we can already expect that these contributions, precisely because they have the same eigenvalue under $\mathfrak{D}_{1,2} $, will be prime candidates for defining an internal (weighted) product, see \ref{subsec: Leading order}.

Moreover, is worth noticing that the superblocks can be consistently decomposed in terms of 1-dimensional conformal blocks $\tilde{g}_\Delta(z) = (-z)^\Delta\, {}_{2}F_{1}(\Delta,\Delta,2\Delta;z) $ as follows
\begin{equation}
    G_\Delta(z) = \tilde{g}_\Delta(z) + \frac{\Delta}{2+2\Delta}\tilde{g}_{\Delta+1}(z) + \frac{\Delta^2}{4(3+4\Delta(2+\Delta))}\tilde{g}_{\Delta+2}(z)\,.
\end{equation}

We move on to the t-channel. In this case we focus on the proximity of the defect insertions at $(23)$ and $(14)$. The latter may initially appear unusual due to the sequential ordering of the correlator insertions with $t_i < t_{i+1}$. However, as discussed in \cite{Liendo:2018ukf}, crossing symmetry allows for the interchange of the second and fourth insertions. We will see this better in the following. For the moment, notice that this implies that the t-channel is equivalent to the previous configuration. Therefore, it is useful to introduce a cross invariant function which can be defined as 
\begin{equation}
\label{fhatdef}
    \hat f(\chi)=\chi^{-2\Delta} f(z)=\chi^{-2\Delta} f\left(\frac{\chi}{\chi - 1}\right),
\end{equation}
with the property 
\begin{equation}
    \hat{f}(\chi) = \hat{f}(1-\chi) .
\end{equation}
This clearly relates the s-channel limit, $\chi\rightarrow 0$, with the t-channel one, $\chi\rightarrow 1$. Thus the $f$-correlator can be written as
\begin{equation}
    \langle \mathbb{R}(t_{1})\bar{\mathbb{R}}(t_{2})\mathbb{R}(t_{3})\bar{\mathbb{R}}(t_{4}) \rangle 
= \frac{C_{\Phi}^{2}}{t_{12}^{2}t_{34}^{2}}f(z)=\frac{C_{\Phi}^{2}}{t_{12}^{2}t_{34}^{2}}\chi^2 \hat{f}(\chi) = \frac{C_{\Phi}^{2}}{t_{13}^{2}t_{24}^{2}}\hat{f}(\chi) ,
\end{equation}
which is explicitly crossing invariant. The $\hat{f}(\chi)$ can also be written in terms of a block expansion and its expression is given by
\begin{equation}
    \hat{f}(\chi) = \frac{1}{\chi^2}+\sum_{\Delta > 0 } a_\Delta\tilde{G}_{\Delta}(\chi),
\end{equation}
with $\tilde{G}_{\Delta}(\chi) = \frac{1}{\chi^2}{G}_{\Delta}(\chi)= \chi^{\Delta-2} {}_{2}F_{1}(\Delta,\Delta+2,2\Delta+2;\chi)$.

\subsubsection{$h$-correlator}
\label{subsec: h correlator}
We start again from the analysis of the s-channel, which is important in order to make contact with the s-channel of the $f$-correlator. Again, one expands around the insertions at $(12)$ and $(34)$, or $\chi\rightarrow 0$. One can, once more, exploit the chiral-antichiral OPE and expect $h(\chi)$ to have the functional expansion
\begin{equation}
\label{h s channel}
    h(\chi) = 1 + \sum_{\Delta}b_{\Delta}
    G_{\Delta}(\chi),
\end{equation}
with $G_\Delta = \chi^{\Delta}\, {}_{2}F_{1}(\Delta,\Delta,2\Delta+2;\chi)$ being obtained by solving \eqref{Ghypergeometric} and by imposing consistency with the OPE expansions. In this respect, the s-channel OPE coefficients of both correlators are specified by the 3-point coefficients
\begin{equation}
    a_\Delta= c_{\mathbb{R}\bar{\mathbb{R}}\mathcal{O}_\Delta} c_{\mathbb{R}\bar{\mathbb{R}}\mathcal{O}_\Delta},\qquad b_\Delta = c_{\mathbb{R}\bar{\mathbb{R}}\mathcal{O}_\Delta} c_{\bar{\mathbb{R}}\mathbb{R}\mathcal{O}_\Delta}, 
\end{equation}
where $\mathcal{O}_\Delta$ denotes the exchanged operator of conformal dimension $\Delta$. As it can be seen by applying a parity transformation \cite{Bianchi:2020hsz}, the two coefficients are related by
\begin{equation}
\label{relation between OPE coefs}
    a_\Delta = (-1)^{s_{\mathcal{O}_\Delta}} b_\Delta,
\end{equation}
where $s_{\mathcal{O}_\Delta}$ stands for the charge of the exchanged operator under parity. Taking, for example, operators schematically defined as
$\mathcal{O}_{2+n} \sim \mathbb{R}\partial_t^n\bar{\mathbb{R}}$, see section \ref{sec:bootstrap}, the corresponding charge is $s_{\mathcal{O}_{2+n}}=n$. In passing, let us also notice that \eqref{relation between OPE coefs} depends solely on the quantum number of the exchanged operator under parity, and thus it remains valid even when perturbations are considered.

Finally, we consider the t-channel. In this last case, one expands around the insertions at $(14)$ and $(24)$, or $\chi\rightarrow 1$, allowing to exploit, for the first time, the chiral-chiral OPE. Let us recall the crossing symmetry briefly discussed for the $f$-correlator: once one identifies the endpoints at infinity, it is possible to swap the positions of points 2 and 4 \cite{Liendo:2018ukf}. This allows to relate the defect insertions
\begin{equation}
\label{crossing symm}
\langle \mathbb{R}(t_{1})\bar{\mathbb{R}}(t_{2})\bar{\mathbb{R}}(t_{3})\mathbb{R}(t_{4}) \rangle =\langle \mathbb{R}(t_{1}){\mathbb{R}}(t_{4})\bar{\mathbb{R}}(t_{3})\bar{\mathbb{R}}(t_{2}) \rangle\;,
\end{equation}
and to exploit the chiral-chiral OPE. In terms of $h(\chi)$, \eqref{crossing symm} implies
\begin{equation}
    (1-\chi)^2 h(\chi) = \chi^2 h(1-\chi)\;.
\end{equation}
It follows that the correlator can be rewritten conveniently as
\begin{equation}
    \langle \mathbb{R}(t_{1})\bar{\mathbb{R}}(t_{2})\bar{\mathbb{R}}(t_{3})\mathbb{R}(t_{4}) \rangle
= \frac{C_{\Phi}^{2}}{t_{14}^{2} t_{23}^{2}}h(1-\chi)= \frac{C_{\Phi}^{2}}{t_{14}^{2} t_{23}^{2}}\hat{h}(\chi) ,
\end{equation}
where we have defined 
\begin{equation}
\label{h hat}
    \hat{h}(\chi) = \Bigl(\frac{1-\chi}{\chi}\Bigr)^{2}h(\chi)\;.
\end{equation}

Now we can perform the CPW expansion and, again, we rely on the analysis of subsection \ref{Selection rules} regarding the chiral-chiral channel. In this case, we have seen that only a single component of each  supermultiplets, with all its conformal family, participates the OPE. It follows that, in this case, the expansion is given in terms of the 1-dimensional conformal blocks \cite{Dolan:2003hv}
\begin{equation}
    g_{\Delta}(\chi) = \chi^{\Delta}\, {}_{2}F_{1}(\Delta,\Delta,2\Delta;\chi).
\end{equation}
If follows that the CPW expansion (in terms of new coefficients $\mathrm{b}_\Delta$) is
\begin{equation}
    \hat{h}(\chi) = \mathrm{b}_{2j_0}g_{2j_0}(1-\chi)+\mathrm{b}_{2j_0+1}g_{2j_0+1}(1-\chi)+\sum_{\Delta>2j_0+1}\mathrm{b}_{\Delta}g_{\Delta}(1-\chi),
\end{equation}
where, as a consequence of the selection rule \eqref{phi-phi selection rule}, the sum is taking into account the protected and unprotected long contributions with $\Delta>2j_0+1$. Specializing to our case $j_0 = 1$, it reads
\begin{equation}
\label{CPW h hat t-channel}
    \hat{h}(\chi) = \mathrm{b}_{2}g_{2}(1-\chi)+\mathrm{b}_{3}g_{3}(1-\chi)+\sum_{\Delta\geq 4}\mathrm{b}_{\Delta}g_{\Delta}(1-\chi).
\end{equation}
\section{Analytic bootstrap}
\label{sec:bootstrap}

At strong coupling, the Wilson lines are mapped through holography to minimal surfaces extending in the AdS bulk and ending along the operator contour on the boundary \cite{Maldacena:1998im}.\footnote{See chapters 12 and 13 of \cite{Drukker:2019bev} for a review of the ABJM case.} The induced metric on the minimal surface is $AdS_2$, which is dual to a CFT$_1$ on the defect \cite{Polchinski_2011,Giombi:2017cqn}. The operators of the superdisplacement multiplet are associated with the fluctuations near the minimal surface of transverse string modes, the leading contribution to a 4-point correlation function arising from a disconnected Witten diagram in $AdS_2$. The first-order correction, on the other hand, is given by the connected 4-point Witten diagram \cite{Giombi:2017cqn,Bianchi:2020hsz}. 

The strong coupling regime can also be accessed by analytic bootstrap methods, which we proceed to perform in this section. The idea is to expand the functions in terms of a small parameter $\epsilon$
\bea
\label{expansionsbootstraofh}
f(z)=f^{(0)}(z)+\epsilon f^{(1)}(z)+O(\epsilon^2),\qquad
h(\chi)=h^{(0)}(\chi)+\epsilon h^{(1)}(\chi)+O(\epsilon^2),
\eea
with the superscripts indicating the leading and  the next-to-leading (NLO) order terms of the expansions, to be eventually matched with the holographic contributions mentioned above.

By general arguments, the parameter $\epsilon$ scales like the inverse string tension $1/T\sim 1/\sqrt{\lambda}$ and is therefore small at strong coupling in the gauge theory, as stated above. The precise mapping depends on the specific ${\cal N}=4$ theory (and, therefore, coupling $\lambda$) one considers and can only be established after an explicit comparison with the Witten diagram computation, not being fixed by symmetry.  

To emphasize the physical meaning of this small parameter, we derive below an explicit relation between $\epsilon$ and $C_\Phi$ (or, equivalently, between $\epsilon$ and the Bremsstrahlung function) following the ideas in  \cite{Drukker:2022pxk,Bliard:2023zpe} based on the computation of the Zamolodchikov metric of the defect conformal manifold.\footnote{We thank Nadav Drukker for suggesting this computation.}


\subsection{Leading order}
\label{subsec: Leading order}

We start from the leading order, which is given by Wick contractions of generalized free fields (GFF)
\cite{Heemskerk:2009pn,Fitzpatrick:2011dm,Fitzpatrick:2012yx,Gaiotto:2013nva,Giombi:2017cqn}. For example, the 4-point correlation function $\langle \mathbb{R}(t_{1})\bar{\mathbb{R}}(t_{2}){\mathbb{R}}(t_{3})\bar{\mathbb{R}}(t_{4}) \rangle$ factorizes in this limit as products of 2-point correlation functions
\begin{equation}
     \langle \mathbb{R}(t_{1})\bar{\mathbb{R}}(t_{2})\rangle \langle{\mathbb{R}}(t_{3})\bar{\mathbb{R}}(t_{4}) \rangle  + \langle \mathbb{R}(t_{1})\bar{\mathbb{R}}(t_{4})\rangle \langle{\bar{\mathbb{R}}}(t_{2})\mathbb{R}(t_{3}) \rangle = C_{\Phi}^2\Bigl(\frac{1}{t_{12}^2t_{34}^2} +\frac{1}{t_{14}^2t_{32}^2}\Bigr)= \frac{C_{\Phi}^2}{t_{12}^2t_{34}^2}(1 +z^2),
\end{equation}
where \eqref{2pnt-components} and the explicit expression \eqref{crossratioszchi} for $z$ have been used. For \eqref{superprimary 4 point correlators}, one obtains 
\begin{equation}
\label{explicit leading order functions}
    f^{(0)}(z) = 1 + z^2,\qquad h^{(0)}(\chi) = 1 + \chi^2\,.
\end{equation}
As expected, the two expressions have the same functional form. In fact, the operators that are exchanged in the s-channels of both 4-point functions of \eqref{superprimary 4 point correlators} are ${\mathbb{R}}\partial_t^n \bar{\mathbb{R}}$, besides the identity, while for the t-channel of the $h$-correlator they are ${\mathbb{R}}\partial_t^n {\mathbb{R}}$. All these have dimensions $\Delta^s_{n}=\Delta^t_{n}= 2+ n$.\footnote{Here and in the following,  $\Delta^t_{n}$ and $\gamma_n^t$ refer exclusively to the t-channel of the $h$-correlator, \textit{i.e.} the chiral-chiral channel. }

To get the conformal data, one has to identify the OPE coefficients of the CPW expansion. By using orthogonality conditions of the conformal blocks in subsection \ref{subsec: Superconformal blocks}, it is a simple task to project out the coefficients (see appendix \ref{app:orthogonality} for details). For our discussion we will only need the relations
\cite{Dolan:2003hv,Bianchi:2020hsz}
\begin{equation}
    \oint \frac{dz}{2\pi i}\omega(z) G_{n+1}(z)G_{-2-m}(z) = \delta_{n,m},
    \qquad \oint \frac{d\chi }{2\pi i}\rho(\chi) g_{n+1}(1-\chi)g_{-m}(1-\chi) = \delta_{n,m}\,,
\end{equation}
where the densities are $\omega(z) = -\frac{1}{(1-z)^2}$ and $\rho(\chi) = -\frac{1}{(1-\chi)^2}$, and the circles close counter-clock-wise around the points $z=0$ and $\chi=1$. Notice that, as discussed in the previous section, the contribution over which we project $G_{m+1}(z) $ is precisely $G_{-2-m}(z)$, {\it i.e.} the shadow one obtains when solving the differential equation. 
Equating the functions \eqref{explicit leading order functions} with the CPW expansion and exploiting the orthogonality, one gets right away the coefficients. For the s-channel, these are
\begin{equation}
    a^{(0)}_{n} = \frac{\sqrt{\pi} 2^{-2(2+n)} (n+1)\Gamma(n+5)}{(n+2)\Gamma(n+\frac{5}{2})},
\end{equation}
and, from the discussion of subsection \ref{subsec: h correlator}, one finds $b^{(0)}_n= (-1)^n\, a_n^{(0)}$. For the t-channel, only even $n$ contribute:
\begin{align}
\mathrm{b}^{(0)}_{n} &=\frac{\sqrt{\pi } 2^{-(1+2n)} (n+1) \Gamma (n+3)}{\Gamma\left(n+\frac{3}{2}\right)},\quad\text{if $n$ even},\cr 
\quad \mathrm{b}^{(0)}_{n} &=0, \;\qquad\qquad\qquad\qquad\qquad\qquad\text{if $n$ odd}.
\end{align}
This is expected,  since only for even $n$ the exchanged operators $\mathbb{R}\partial_t^n {\mathbb{R}}$ respect the $\mathbb{Z}_{2}$ symmetry  $t\rightarrow -t$.


\subsection{Next-to-leading order}
\label{subsec: Next-to-leading order}

We move now to investigate the NLO of the expansions \eqref{expansionsbootstraofh}. Following \cite{Bianchi:2020hsz}, one can start from an ansatz for $f^{(1)}$ and $h^{(1)}$. It is however particularly convenient to focus on $\hat{f}(\chi)$ in \eqref{fhatdef}, since one can exploit crossing invariance to fix its generic structure, given by
\begin{equation}
\label{f hat ansatz}
    \hat{f}^{(1)}(\chi) = r(1-\chi)\log(\chi)+r(\chi)\log(1-\chi)+q(\chi),
\end{equation}
where $q(\chi)$ and $r(\chi)$ are rational functions expanded as \cite{Bianchi:2020hsz} 
\begin{equation}
\label{rational functions}
    r(\chi) = \sum_k r_k \chi^k \qquad q(\chi)=\sum_l q_l \chi^l (1-\chi)^l.
\end{equation}
In the following, we will investigate how to fix the coefficients $r_k$ and $q_l$. Plugging the ansatz in $f(z)$, one has
\begin{equation}
\label{f ansatz}
    {f}^{(1)}(z) = \frac{z^2}{(z-1)^2}\Bigr[r\Bigl( \frac{1}{1-z}\Bigr)\log(-z)-\Bigl(r\Bigl( \frac{1}{1-z}\Bigr)+r\Bigl( \frac{z}{z-1}\Bigr)\Bigr)\log(1-z)+q\Bigl( \frac{z}{z-1}\Bigr)\Bigr].
\end{equation}
Similarly, one can find an analogous expression for $h^{(1)}$ by recalling that in the s-channels the two functions have the same functional form. Hence, by replacing $z \rightarrow \chi$, one obtains the expression for $h^{(1)}(\chi)$ , up to the sign in the logarithm\footnote{The sign is changed in order to require consistency with the first-order expansion of the expression \eqref{h s channel}, see also \cite{Bianchi:2020hsz}.  By doing the expansion explicitly two contributions given by $\log(\chi)$ and $\log(1-\chi)$ factorize. Equivalently, one can consider absolute values in the arguments as done in \cite{Giombi:2017cqn}.}
\begin{equation}
\label{h ansatz}
    h^{(1)}(\chi) = \frac{\chi^2}{(\chi-1)^2}\Bigr[r\Bigl( \frac{1}{1-\chi}\Bigr)\log(\chi)-\Bigl(r\Bigl( \frac{1}{1-\chi}\Bigr)+r\Bigl( \frac{\chi}{\chi-1}\Bigr)\Bigr)\log(1-\chi)+q\Bigl( \frac{\chi}{\chi-1}\Bigr)\Bigr],
\end{equation}
and, plugging \eqref{h ansatz} in \eqref{h hat}, also the expression for $\hat h^{(1)}(\chi)$
\begin{equation}
\label{h hat ansatz}
     \hat{h}^{(1)}(\chi) = r\Bigl( \frac{1}{1-\chi}\Bigr)\log(\chi)-\Bigl(r\Bigl( \frac{1}{1-\chi}\Bigr)+r\Bigl( \frac{\chi}{\chi-1}\Bigr)\Bigr)\log(1-\chi)+q\Bigl( \frac{\chi}{\chi-1}\Bigr).
\end{equation}

These must match with the perturbed CPW expansions. In particular, we expect the conformal dimensions to become anomalous
\bea
\Delta^{s}_n = 2 + n + \epsilon \gamma^{s}_n, 
\qquad
\Delta^{t}_n = 2 + n + \epsilon \gamma^{t}_n,
\eea
for the s- and t-channel, respectively. Similarly, the CPW coefficients will now be given by 
\bea
a_n = a^{(0)}_n + \epsilon a^{(1)}_n, \qquad \text{b}_n = \text{b}^{(0)}_n + \epsilon \text{b}^{(1)}_n .
\eea
The NLO expansions for the functions \eqref{CPW f s-channel} and \eqref{CPW h hat t-channel} are then explicitly given by
\begin{align}
    {f}^{(1)}(z) &= \sum_{n\geq 0}(-z)^{n+2}\Bigl( a^{(1)}_{n}F_{2+n}(z)+ \gamma^s_{n} a^{(0)}_{n}F_{2+n}(z)\log(-z)+ \gamma^s_{n} a^{(0)}_{n}\partial_\Delta F_{\Delta}(z)\vert_{2+n}\Bigr),\cr
    \hat{h}^{(1)}(\chi) &= \sum_{n\geq 0}(1-\chi)^{n+2}\Bigl( \mathrm{b}^{(1)}_{n}\tilde{F}_{2+n}(1-\chi)
    \cr & \hskip 4cm + \gamma^t_{n} \mathrm{b}^{(0)}_{n}\tilde{F}_{2+n}(1-\chi)\log(1-\chi)+ \gamma^t_{n} \mathrm{b}^{(0)}_{n}\partial_\Delta \tilde{F}_{\Delta}(1-\chi)\vert_{2+n}\Bigr),\cr &
    \label{h hat expansion}
\end{align}
with simplified notation: $F_\Delta(z) = {}_2 F_{1}(\Delta,\Delta,2\Delta+2;z)$ and $\tilde{F}_\Delta(1-\chi) = {}_2 F_{1}(\Delta,\Delta,2\Delta;1-\chi)$.
One may infer the CFT data as it was done in the previous section at leading order. In particular, the anomalous dimensions can be extracted and related to the ansatz by 
\begin{align}
    \gamma_n^s =&\frac{1}{a^{(0)}_n}\oint \frac{dz}{2\pi i}\omega(z) \Bigl[\frac{z^2}{(z-1)^2} r\Bigl(\frac{1}{1-z}\Bigr) \Bigr]G_{-3-n}(z),\cr
   \gamma_n^t =& -\frac{1}{\mathrm{b}^{(0)}_n}\oint \frac{d\chi }{2\pi i}\rho(\chi) \Bigl[r\Bigl( \frac{1}{1-\chi}\Bigr)+r\Bigl( \frac{\chi}{\chi-1}\Bigr)\Bigr] g_{-1-n}(1-\chi),\qquad\text{for $n$ even} \,,\label{generic gamma t}
\end{align}
while $\gamma_n^t=0$ for $n$ odd. Here a comment is in order: from the evaluation of the anomalous dimensions through this procedure, one is not extracting the contributions of individual operators entering the OPE, but rather a linear combination thereof, weighted by the corresponding CPW coefficients \cite{Bianchi_2017,Giombi:2017cqn}. The unmixing of these contributions is a highly nontrivial task \cite{Ferrero:2021bsb,Ferrero:2023gnu,Ferrero:2023znz}, which we do not address here.

To fix the coefficients in \eqref{rational functions}, one can first truncate their expansions borrowing some arguments from \cite{Fitzpatrick:2010zm,Polchinski_2011,Giombi:2017cqn,Gimenez-Grau:2019hez, Bianchi:2020hsz,Alday:2014tsa,Heemskerk:2009pn} regarding the large-$n$ behaviour of the anomalous dimensions. The lore goes as follow: the increase in anomalous dimensions is tied to the local interactions occurring within the AdS dual counterpart. As the relevance of these interactions decreases, the growth in anomalous dimensions becomes more significant. In aiming to bootstrap a leading correction for the holographic correlator, it is sensible to prioritize solutions characterized by minimal, or mildest, growth. In the CFT this can be justified a posteriori knowing that $\gamma_n<0$, so that the mildest growth of the anomalous dimension is the one that guarantees reliability for the largest span of values of $n$ \cite{Giombi:2017cqn}. If $\gamma_n$ grows too fast, the contribution $\epsilon\gamma_n^{s,t}$ will grow faster compared to the leading order $2+n$ until, eventually, hitting the unitarity bound. Through the prescription \eqref{generic gamma t} one can compute the anomalous dimensions exploiting the rational function $r(\chi)$. This leads to the parametric expressions
\begin{align}
    \gamma^s_n &= \frac{1}{4}(1+n)(4+n)r_{-3}+r_{-2}+\frac{2+(-1)^n(6+5n+n^2)}{2(4+5n+n^2)}r_{-1}-\frac{(-1)^n}{6}(6+5n+n^2)r_{0}+O(n^4),\cr
    \gamma^t_n &= \frac{1}{4}n(n+3)r_{-3}-r_{-2}+\frac{1}{(1+n)(2+n)}r_{-1}+O(n^4),\qquad\text{for $n$ even}.
\end{align}
To meet the mildest-$n$ behaviour, the coefficients of the expansion of $r(\chi)$ must be  $r_k\neq0$ only for $-3\leq k\leq 0$. However, there are some oscillating contributions appearing on the s-channel anomalous dimension. 
Motivated by the results in \cite{Giombi:2017cqn, Gimenez-Grau:2019hez, Bianchi:2020hsz}, one could think to further fix them by expecting an universal strong coupling trend for $n\gg 1$ and $\epsilon\ll 1$ but with $n\epsilon$ fixed, see {\it e.g.} \cite{Giombi:2017cqn}. This consideration would enable to fix $r_0=0$ (which will also emerge from requiring the regularity of the ansatz anyway), but it does not allow to draw any conclusion on the $r_{-1}$ coefficient, as it become negligible at large $n$. Crucially, the requirement that $\gamma^t_0 = 0$, following from the selection rules, introduces a further condition that must be satisfied, and that is
\begin{equation}
    2r_{-2}-r_{-1}=0.
\end{equation}
This constraint and $r_0=0$ can be equivalently obtained by requiring that the series expansion around $\chi\rightarrow 1$ of the ansatz \eqref{h hat ansatz} starts at $(1-\chi)^{4}$, precisely as in \eqref{h hat expansion}.

As anticipated, other restrictions can be drawn from imposing the condition of regularity for the ansatz \eqref{f hat ansatz} and \eqref{h hat ansatz}. This translates into the requirement of pole cancellations in the regimes $\chi\rightarrow 1$ (or equivalently $\chi\rightarrow 0$) for $\hat{f}^{(1)}(\chi)$  and $\chi\rightarrow 1$ for $\hat{h}^{(1)}(\chi)$. One obtains the conditions
\begin{equation}
    q_{-2}-r_{-3}=0, \qquad 2q_{-2}+q_{-1}-\frac{r_{-3}}{2}-r_{-2}=0,
\end{equation}
and all others $q_l=0$. All other constraints that can be obtained from similar analyses will just be redundant.

Summarizing, one can consider the ansatz specified by the rational functions
\begin{equation}
    r(\chi) = \frac{r_{-3}}{\chi^3}+\frac{r_{-2}}{\chi^2}+\frac{r_{-1}}{\chi},\qquad 
    q(\chi) = \frac{q_{-2}}{\chi^2(1-\chi)^2}+\frac{q_{-1}}{\chi(1-\chi)},
\end{equation}
with the five coefficients subject to
\begin{equation}
    r_{-3}=q_{-2},\qquad r_{-2}=\frac{3 q_{-2}}{2}+q_{-1},\qquad r_{-1}=3q_{-2}+2q_{-1}.
\end{equation}
There remain then two independent coefficients, which are not fixed by internal consistency, crossing symmetry, and so on. This is not an unexpected result, as it can justified by drawing  analogies from cases in $4$ dimensions. In particular, for $\mathcal{N}=4$ super Yang-Mills the corresponding solutions are defined up to an overall parameter \cite{Giombi:2017cqn}. However, when one studies less supersymmetric cases, {\it e.g.} $\mathcal{N}=2$, one finds that the solutions are expressed in terms of two free parameters \cite{Gimenez-Grau:2019hez}. Something similar happens for 3-dimensional theories. The solutions found for the ABJM case \cite{Bianchi:2020hsz} were defined up to a single overall parameter too, while here we find, as for $\mathcal{N}=2$, a 2-parameter family of solutions. 

In order to state the final answer one can proceed by absorbing one of the coefficient in the expansion parameter $\epsilon$ and, in analogy with the $SU(1,1\vert 3)$ case \cite{Bianchi:2020hsz}, we also change the overall sign.\footnote{\label{footnotesign} In fact, the discussion in this section is true regardless of the overall sign of the ansatz. This sign change was justified in \cite{Bianchi:2020hsz} by the comparison with the holographic dual.} Hence, by reabsorbing $q_{-2}$ into the definitions of $\epsilon$ and expressing the ratio of the two free parameters as 
\bea
\xi \equiv  \frac{q_{-1}}{q_{-2}}, 
\eea
\eqref{f hat ansatz} now reads
\begin{align}
    \hat{f}^{(1)}(\chi) =& -\left(\frac{2 \xi +3}{\chi }+\frac{\xi +\frac{3}{2}}{\chi ^2}+\frac{1}{\chi ^3}\right) \log (1-\chi )\cr & -\left(\frac{2 \xi +3}{1-\chi }+\frac{\xi +\frac{3}{2}}{(1-\chi )^2}+\frac{1}{(1-\chi )^3}\right) \log (\chi )-\frac{\xi }{(1-\chi ) \chi }-\frac{1}{(1-\chi )^2 \chi ^2}.
    \label{f hat xi}
\end{align}
The expression for $f^{(1)}(z)$ is given instead by
\begin{align}
f^{(1)}(z)=& -1+(2+\xi)z-z^2-\frac{(1-z)^2\left(2 z^2-(2 \xi +1) z+2\right) \log (1-z)}{2 z}\cr
&-\frac{z^2 \left(6 \xi +2 z^2-(2 \xi +7) z+11\right) \log (-z)}{2 (1-z)}.
\label{f xi}
\end{align}
As before,  $h^{(1)}(\chi)$ can be obtained from  $f^{(1)}(z)$ through the usual mapping, while for \eqref{h hat ansatz} one has explicitly
\bea
    \hat{h}^{(1)}(\chi)&=&
    \frac{\chi-1}{2\chi^3}
    \Bigl[ 2(\chi-1)\chi(1-(2+\xi)\chi+\chi^2)+ (\chi-1)^3(2-(1+2\xi)\chi +2\chi^2 )\log(1-\chi)\cr & & \hskip 2cm +\chi^3(2 \xi  (\chi -3)-2 \chi ^2+7 \chi -11)\log(\chi) \Bigr].
\eea
With these parameterizations, the anomalous dimensions for the two channels are now expressed as
\begin{eqnarray}
\label{resultgammasandt}
\gamma^{s}_n &=&-\frac{4+n(n+1)\left[2+n+n^2 +4\xi+2(-1)^n (3+2 \xi)\right]}{4 \left(-2+n+n^2\right)},\cr
\gamma^{t}_n &=&-\frac{1}{4}n(n+3)+\frac{n(3+n)(3+2\xi)}{2(1+n)(2+n)},\qquad\text{for $n$ even},
\end{eqnarray}
and the NLO OPE coefficients can be consistently written as \cite{Heemskerk:2009pn,Fitzpatrick:2011dm,Alday:2014tsa}
\begin{equation}
      a^{(1)}_n=\frac{\partial}{\partial n}\Bigl(a^{(0)}_n\gamma^s_n\Bigr),
\end{equation}
and
\begin{equation}
  \mathrm{b}^{(1)}_n=\frac{\partial}{\partial n}\Bigl(\mathrm{b}^{(0)}_n\gamma^t_n\Bigr),
    \qquad\text{for $n$ even}.
\end{equation}


\subsection{Relation to the 1/2 BPS line in ABJM}
\label{subsec: Relations to ABJM}

By looking at most results above, {\it e.g.} \eqref{f hat xi} for $\hat f^{(1)}(\chi)$ or \eqref{resultgammasandt} for the anomalous dimensions, one notices that $\xi$ screams to be set to $\xi = -3/2$, to drastically simplify those expressions. However, as discussed, this cannot be done by invoking internal consistency or crossing symmetry. The presence of the free parameter $\xi$ is simply due to the ${\cal N}=4$ case being less constrained by supersymmetry than ABJM. 

We can however make contact with the ABJM case\footnote{We specialize to those configurations that can be directly derived from the ABJM theory, for example from quotients of the latter \cite{Drukker:2019bev}.} by identifying the operators in the ABJM superdisplacement multiplet which map to the ones in \eqref{superdisplacementcharges}, which is
\begin{equation}
    L\bar{A}[{\ft{3}{2}}]^{(0,0)}_{\frac{1}{2}}\rightarrow L\bar{A}[1]^{(0)}_{1}.
\end{equation}
In ABJM, the primary operator is a fermion with $\Delta=1/2$, $j_0=3/2$ and singlet under R-symmetry.
We may then compare the 4-point functions of the defect operators with the same dimension $\Delta$ in ABJM and in ${\cal N}=4$. In the comparison, one has to take into account that the normalizations of the 2-point functions and the small parameters of the expansions are different in the two cases, so we call them $C_\textrm{ABJM}$ and $\epsilon_\textrm{ABJM}$ in the ABJM case, to distinguish them from $C_\Phi$ and $\epsilon$ above.

The displacement operator $\mathbb{D}$, which appears in both theories and is neutral under the preserved R-charges, is the natural place to start the comparison. Its 4-point correlation function for the ABJM case has been obtained in \cite{Bianchi:2020hsz} and is a rather complicated expression
\begin{align}
\langle\mathbb{D}(t_1)\bar{\mathbb{D}}(t_2)\mathbb{D}(t_3)\bar{\mathbb{D}}(t_4)\rangle_{\text{ABJM}}=&
\frac{(12 C_{\text{ABJM}})^2}{t_{12}^4 t_{34}^4}\frac{1}{36}
\Big[36 \text{f}-36 (z^4+z) \text{f}' +18 z^2 (-14 z^3+3 z^2+1) \text{f}''\nonumber\\
& \hskip 3cm -6 z^3 \left(55 z^3-39 z^2+3 z+1\right) \text{f}^{(3)}\cr & \hskip 3cm -3 z^4 \left(46 z^3-63 z^2+18 z-1\right) \text{f}^{(4)}\nonumber\\
& \hskip 3cm -3 (z-1)^2 z^5 (7 z-1) \text{f}^{(5)} -(z-1)^3 z^6 \text{f}^{(6)}
 \,\Big]\cr & 
\label{displacement f exp ABJM}
\end{align}
of the function $\text{f}(z)=f_\textrm{ABJM}(z)$ and its first six derivatives. 
 
At leading order one finds $\text{f}^{(0)}(z)=1-z$ \cite{Bianchi:2020hsz}. Notice that since it encodes the information of the superprimary 4-point function of the ABJM 1/2-BPS Wilson line, it has clearly a different functional form compared to the ${\cal N}=4$ case in \eqref{explicit leading order functions}. 
Nonetheless, when computing the 4-point function at this order, one obtains  
\begin{equation}
    \langle\mathbb{D}(t_1)\bar{\mathbb{D}}(t_2)\mathbb{D}(t_3)\bar{\mathbb{D}}(t_4)\rangle_{\text{ABJM}} = \frac{(12 C_{\text{ABJM}})^2}{t_{12}^4 t_{34}^4}\left( 1+ z^4\right)+O(\epsilon_\textrm{ABJM}),
\end{equation}
which is, upon the overall factor, precisely the same functional form that one gets in ${\cal N}=4$, see \eqref{displacement 4 point function} with $f(z)=f^{(0)}(z)$ in \eqref{explicit leading order functions}. The NLO result for ABJM was obtained in \cite{Bianchi:2020hsz} and reads
\begin{equation}
     f_\textrm{ABJM}^{(1)}(z)=z-1 - \frac{(1-z)^3}{z}\log(1-z)+z(3-z)\log(-z).
\end{equation}
When plugged into \eqref{displacement f exp ABJM}, the 4-point function of the displacement operator of the ABJM theory becomes
\begin{align}
\langle\mathbb{D}(t_1)\bar{\mathbb{D}}(t_2)\mathbb{D}(t_3)\bar{\mathbb{D}}(t_4)\rangle_{\text{ABJM}}=&
\frac{(12 C_{\text{ABJM}})^2}{t_{12}^4 t_{34}^4}\, 
\Bigl[1+z^4+2\epsilon_{\text{ABJM}}\Bigl(-8 -z -\frac{7z^2}{6} -z^3-8z^4
\cr
&+\Bigl(3 - \frac{8}{z}+3 z^4 -8z^5\Bigr)\log(1-z) + z^4 (8z-3)\log(-z)\Bigr)\Bigr].
\label{4pnt displacement leading and NLO order}
\end{align}
Remarkably, if we evaluate \eqref{displacement 4 point function} for $\xi=-3/2$, we precisely reproduce the formula above for ABJM.

For reference, the anomalous dimensions and coefficients of the CPW expansion for $\xi=-3/2$ reduce to
\begin{align}
\gamma^{s}_n =&-n^2 -5n-4,\cr
    \gamma^{t}_n=&-n^2-3n,\qquad\text{for $n$ even},
\end{align}
and
\begin{align}
    a^{(1)}_n &= a^{(0)}_n\Bigl[ -2n-5 + \gamma^s_n \Bigl( \psi(n+5)-\psi\left(n+\ft{5}{2}\right)-2\log 2 +\frac{1}{(n+1)(n+2)} \Bigr) \Bigr],\cr
    \mathrm{b}^{(1)}_n &= \mathrm{b}^{(0)}_n\Bigl[ -2n-3 + \gamma^t_n \Bigl( \psi(n+3)-\psi\left(n+\ft{3}{2}\right)-2\log 2+\frac{1}{n+1} \Bigr) \Bigr],\qquad\text{for $n$ even},\cr &
\end{align}
where $\psi(z)=\Gamma'(z)/\Gamma(z)$ is the digamma function.


\subsection{Relation to the 1/2 BPS line in $\mathcal{N}=2$ theories}

One can exploit what has been computed so far to guess what the result for the bootstrap of the displacement operator would be like for the 1/2 BPS Wilson line defect of $\mathcal{N}=2$ SCSM \cite{Gaiotto:2007qi}. To do this, one needs to recall the different superdisplacement multiplets in the 3-dimensional $\cN=2,4,6$ cases~\cite{Agmon:2020pde}: 
\begin{equation}
\begin{array}{lllll}
\cN=6:\qquad & [\ft{3}{2}]^{(0,0)}_{\frac{1}{2}} \longrightarrow &  [2]^{(1,0)}_{1} \longrightarrow &  \left[\ft{5}{2}\right]_{\frac{3}{2}}^{(0,1)} \longrightarrow &
  [3]_{2}^{(0,0)}\; , \\
  &&&& \\
\cN =4:\qquad & & [1]_{1}^{(0)} \longrightarrow & \left[1\right]_{\frac{3}{2}}^{(1)} \longrightarrow & [1]_{2}^{(0)}\; , \\
&&&& \\
\cN=2: \qquad &&&  \left[-\ft{3}{2}\right]_{\frac{3}{2}} \longrightarrow &  [-1]_{2}\; ,
\end{array}
\end{equation}
with the usual notation for the charges $[j_0]_{\Delta}^{(\mathcal{R})}$, with $(\mathcal{R})$ standing for the Dynkin labels of the considered representation.

All displacement operators are singlets of the preserved R-symmetry (when present), and the 4-point function is neutral under the Abelian symmetry. Therefore, one expects the same relation to take place, where now the less supersymmetric one is the defect theory of the 1/2 BPS Wilson line in $\cN=2$ and the $\cN=4$ defect plays the same role as the ABJM one in the previous section. Thus, one can expect that whatever function is relative to the superprimary, it should be consistent, upon normalization, with the ${\cal N}=4$ one, precisely as before. Moreover, one can extract much more information about the unknown function.
Notably, as illustrated in the scheme above, the multiplets shorten as the bulk supersymmetry decreases, as expected. Consequently, the superprimary of the less supersymmetric multiplet can be matched with the first $Q$-descendant of the higher supersymmetric theory.
This is true and can be verified for the matching between the ${\cal N}=4$ case and ABJM, as seen above. By exploiting \eqref{fermionic 4 point function}, one can do the same for $\cN=4$  and   $\cN=2$, arriving at the $f$-function for the ${\cal N}=2$ case, both at the leading order
\begin{equation}
    f_{{\cal N}=2}^{(0)}(z) = 1 - z^3,
\end{equation}
and at the NLO
\begin{equation}
   f_{{\cal N}=2}^{(1)}(z) = -9+\frac{z}{2}-\frac{z^2}{2}+9z^3 +\Bigl( 5-\frac{9}{z}+5z^3-9z^4\Bigr)\log(1-z) +(5-9z)z^3\log(-z).
\end{equation}
One can straightforwardly extend this analysis to the $h$-function, the anomalous dimensions, and so on.


\subsection{Comments on the holographic dual}

A natural question at this point is of course to try to  reproduce the results obtained here from the holographic dual \cite{Lietti:2017gtc}, by computing explicitly the disconnected and connected Witten diagrams in $AdS_2$, as pioneered in \cite{Giombi:2017cqn}. In fact, this is not strictly necessary, as one can simply reuse the analysis in \cite{Bianchi:2020hsz} for the ABJM case.

The supergravity dual is of course different in the ${\cal N}=4$ case, namely ${AdS}_4 \times S^7/\mathbb{Q}$, 
where $\mathbb{Q}$ is some appropriate quotient of $S^7$, for example $\mathbb{Q}=(\mathbb{Z}_p\oplus \mathbb{Z}_q)/\mathbb{Z}_k$ for quivers coupled to $p$ hypers and $q$ twisted hypers. This implies that the Kaluza-Klein reduction of the supergravity fields in the internal space is going to be different, {\it e.g.} one does not expect the three massless fields found in \cite{Bianchi:2020hsz} due the $SU(3)_R$ R-symmetry. However, the analysis for the $AdS_2$ fluctuations is going to be the same, correspondingly to the fact that the displacement operator $\mathbb{D}$ is the same as in that case. The Witten diagram for this field is decoupled from the rest, so that the result in \cite{Bianchi:2020hsz} also applies here, consistently with the discussion above in subsection \ref{subsec: Relations to ABJM}.


\subsection{Relation between $\epsilon$ and $C_\Phi$}

It is possible to establish a relation between the expansion parameter $\epsilon$ and the normalization of the 2-point correlation functions $C_\Phi$ or, equivalently, the Bremsstrahlung function of the theory. This was first proposed in \cite{Drukker:2022pxk} and then extended to higher orders at strong coupling in \cite{Bliard:2023zpe}.

The idea is to consider the defect conformal manifold associated with exactly marginal deformations of our 1/2 BPS operators. This manifold is $\mathbb{CP}^1$
with Zamolodchikov metric given by
\begin{equation}
    g_{z\bar{z}}=\langle \mathbb{R}(0)\bar{\mathbb{R}}(1)\rangle = C_{\Phi} \delta_{z\bar{z}}\,,
\end{equation}
which is conformally related to the line element
\begin{equation}
    ds^2_{\mathbb{CP}^1} = \frac{4C_{\Phi}}{(1+\vert z\vert^2)^2}dzd\bar{z}.
\end{equation}
The corresponding Riemann tensor is
\begin{equation}
\label{riemann}
    R_{z\bar{z}z\bar{z}} = 
    -\frac{1}{2}g_{{z\bar{z}}}g_{{z\bar{z}}}\mathcal{R}_{\mathbb{CP}^{1}},\qquad\, R_{z z\bar{z} \bar{z}} =0,
\end{equation}
with Ricci scalar  
\begin{equation}
    \mathcal{R}_{\mathbb{CP}^{1}}=\frac{2}{C_{\Phi}}.
\end{equation}

Now, as suggested in \cite{Drukker:2022pxk}, the Riemann tensor can be also interpreted as the 4-point correlation functions of the primary operator, by extending the Zamolodchikov metric beyond the flat space approximation. There are the two possible orderings seen above, so that one has to consider 
\begin{align}
    t^2_{12}t^2_{34}\langle \mathbb{R}(t_{1})\bar{\mathbb{R}}(t_{2})\mathbb{R}(t_{3})\bar{\mathbb{R}}(t_{4}) \rangle & = g_{z\bar{z}}g_{z\bar{z}} f(z)=C_{\Phi}^2 f(z), \cr
    t^2_{12}t^2_{34}\langle \mathbb{R}(t_{1})\bar{\mathbb{R}}(t_{2})\bar{\mathbb{R}}(t_{3})\mathbb{R}(t_{4}) \rangle & = g_{z\bar{z}}g_{\bar{z}z} h(\chi)=C_{\Phi}^2 h(\chi).
\end{align}
Following the analysis in \cite{Drukker:2022pxk} and identifying our functions with their functions as  $f=2K_1=-2K_2$ and $h=2H_1=-2H_2$, one finds $R_{z z \bar{z}\bar{z}}=0$, consistently with \eqref{riemann} above, and $ R_{z \bar{z}z \bar{z}}=2 g_{z\bar{z}}g_{z\bar{z}}\mathcal{R}$ with
\begin{equation}
    \mathcal{R}=\int_{0}^{1}\frac{d\chi}{\chi^2}\left[\log(1-\chi)\left(\,2h(\chi)-\frac{1}{2}f\left(\frac{\chi}{\chi-1}\right)\right) - \log\chi \left(\,h(\chi)+\frac{1}{2}f\left(\frac{\chi}{\chi-1}\right)\right)\right].
\end{equation}
Evaluating this with $f^{(1)}$ and $h^{(1)}$, one obtains
\begin{equation}
    \mathcal{R} = 2\pi^2 \epsilon.
\end{equation}
Comparing with the Riemann tensor in \eqref{riemann}, one finds the desired relation
\begin{equation}
\epsilon = -\frac{1}{4\pi^2 C_{\Phi}},
\end{equation}
with the minus sign being consistent with having absorbed another minus sign in $\epsilon$ in the Ansatz (see discussion in footnote \ref{footnotesign}). 

If one knows the Bremsstrahlung function of the theory at strong coupling, this relation can be used to determine $\epsilon$ in a way which is independent of the comparison with the holographic computation in terms of Witten diagrams. This was done in  \cite{Drukker:2022pxk} for the 1/2 BPS Wilson loops of ${\cal N}=4$ super Yang-Mills and of the ABJM theory. Unfortunately, such computation of the Bremsstrahlung function has not been performed yet for ${\cal N}=4$ Chern-Simons-matter theories, so that one cannot get the precise expression of $\epsilon$. However, the leading term of the Bremsstrahlung function scales generically like $\sqrt{\lambda}$, so that $\epsilon\sim 1/\sqrt{\lambda}$, as stated at the beginning of this section. 


\section*{Acknowledgements}

We are indebted to Gabriel Nagaoka for collaboration at the initial stages of this work and to Lorenzo Bianchi, Nadav Drukker and Carlo Meneghelli for discussions and suggestions. DT is supported in part by the INFN grant {\it Gauge and String Theory (GAST)} and would like to thank FAPESP’s partial support through the grant 2019/21281-4.
\appendix

\section{Symmetries of the bulk and defect theories}
\label{app:algebra}

We collect here details about our conventions and the symmetries preserved by the bulk ${\cal N}=4$ theory and its 1/2 BPS Wilson lines. We work in Euclidean space $\mathbb{R}^3$. 

The 3-dimensional $\cN=4$ superconformal algebra is $\mathfrak{osp}(4|4)$. Its bosonic subalgebra consists of the 3-dimensional conformal algebra $\mathfrak{so}(1,4)$ and of the R-symmetry algebra $\mathfrak{so}(4)_{R}\simeq \mathfrak{su}(2)_A\oplus\mathfrak{su}(2)_B$. The conformal generators are the rotations $M^{\mu\nu}$, the translations $P^\mu$, the special conformal transformations $K^\mu$ and the dilations $D$, with $\mu,\nu=0,1,2$ and algebra given by
\bea
&&[M^{\mu\nu},M^{\rho\sigma}]=\delta^{\sigma\mu}M^{\nu\rho}-\delta^{\sigma\nu}M^{\mu\rho}+\delta^{\rho\nu}M^{\mu\sigma}-\delta^{\rho\mu}M^{\nu\sigma},\cr
&&[P^\mu,M^{\nu\rho}]=\delta^{\mu\nu}P^{\rho}-\delta^{\mu\rho}P^{\nu},\cr
&&[K^\mu,M^{\nu\rho}]=\delta^{\mu\nu}K^{\rho}-\delta^{\mu\rho}K^{\nu},\cr
&&[P^\mu,K^\nu] = -2\delta^{\mu\nu}D-2 M^{\mu\nu},\cr
&&[D,P^\mu]=P^\mu,\qquad [D,K^\mu]=-K^\mu.
\eea
The R-symmetry generators are $R_{IJ}=-R_{JI}$, with $I,J=1,2,3,4$, and obey
\bea
[R_{IJ},R_{KL}] = \delta_{K[I} R_{J]L}+\delta_{L[J} R_{I]K}.
\eea

The fermionic generators $Q_{I\alpha}$ and $S^\alpha_I$, with spinorial indices $\alpha=\pm$, satisfy
\bea
\label{fermionic-anticom}
&& \{Q_{I\alpha}, Q_{J\beta}\} = 2i\delta_{IJ} (\gamma^\mu)_{\alpha\beta} P_\mu, \cr
&& \{S_I^\alpha, S_J^\beta\} = 2i\delta_{IJ} (\gamma^\mu)^{\alpha\beta} K_\mu, \cr
&& \{Q_{I\alpha}, S_J^\beta\} = \delta_{IJ} \left((\gamma^{\mu\nu})_\alpha{}^\beta M_{\mu\nu} + 2\delta_\alpha^\beta  D\right) + 2\delta_\alpha^\beta R_{IJ}, 
\eea
with $(\gamma^{\mu})_\alpha{}^\beta$ being the Pauli matrices satisfying $\{\gamma^\mu,\gamma^\nu\}=2\delta^{\mu\nu}$ and $\gamma^{\mu\nu}= \frac{1}{2}(\gamma^\mu\gamma^\nu-\gamma^\nu\gamma^\mu)=i\epsilon^{\mu\nu\rho}\gamma^\rho$.

The remaining commutation relations are
\bea
[D,Q_{I\alpha}]=\frac{1}{2} Q_{I\alpha},\qquad &&\qquad [D,{S_I}^\alpha]=-\frac{1}{2}{S_I}^\alpha, \cr
[M^{\mu\nu},Q_{I\alpha}]=-\frac{1}{2}(\gamma^{\mu\nu})_\alpha^\beta Q_{I\beta}, \quad \quad && \quad \quad [M^{\mu\nu},{S_I}^\alpha]=\frac{1}{2}(\gamma^{\mu\nu})^\alpha_{\,\beta} S_I^\beta,\cr
[K^{\mu},Q_{I\alpha}]=i(\gamma^{\mu})_{\alpha\beta} {S_I}^\beta, \quad \quad && \quad \quad [P^{\mu},{S_I}^\alpha]=-i(\gamma^{\mu})^{\alpha\beta} Q_{I\beta}\cr
[R_{IJ},Q_{K\alpha}] = \delta_{IK}Q_{J\alpha}-\delta_{JK}Q_{I\alpha},\quad\quad && \quad\quad [R_{IJ},S_{K}^\alpha] = \delta_{IK}S_{J}^\alpha-\delta_{JK}S_{I}^\alpha,
\eea
where spinorial indexes are raised/lowered with $\epsilon^{\alpha\beta}$ and $\epsilon_{\alpha\beta}$, such that $ \epsilon^{+-}=\epsilon_{-+} = 1$. 

Since the Wilson line insertion breaks half of the supersymmetry, it is convenient to decompose the R-symmetry in terms of the isomorphic $\mathfrak{su}(2)_L\oplus \mathfrak{su}(2)_R$
\begin{equation}
    [R_a{}^b, R_c{}^d] = \delta_{c}^{b} R_a{}^d - \delta_{a}^{d} R_c{}^b, \qquad\qquad  [R^{\dot{a}}{}_{\dot{b}}, R^{\dot{c}}{}_{\dot{d}}] = \delta^{\dot{c}}_{\dot{b}} R^{\dot{a}}{}_{\dot{d}}-\delta^{\dot{a}}_{\dot{d}} R^{\dot{c}}{}_{\dot{b}},
\end{equation}
where $R_{a}{}^{b} = -\frac{1}{4}(\sigma^I\Bar{\sigma}^J)_a^{\;b} R_{IJ}$ and ${\Bar{R}}^{\dot{b}}_{\;\dot{a}}= -\frac{1}{4}(\Bar{\sigma}^I\sigma^J)_{\dot{a}}^{\dot{b}} R_{IJ}$, with  indices $a,b=1,2$ and $\dot a, \dot b=\dot 1,\dot 2$. We take $\sigma^I_{a\dot{a}}=(\mathbb{1},i\sigma^1,i\sigma^2,i\sigma^3)$ and $\Bar{\sigma}^{I\dot{a}a}=(\mathbb{1},-i\sigma^1,-i\sigma^2,-i\sigma^3)$. 

For the fermionic charges we define instead $Q_{\alpha a\dot{a}}=\sigma^I_{a\dot{a}} Q_{I\alpha}$ and similarly for the superconformal charges. Correspondingly, \eqref{fermionic-anticom} becomes
\begin{alignat}{3}
\label{algebra-app}
	\{Q_{\alpha a\dot{a}} , Q_{\beta b\dot{b}}\} &= 2i\varepsilon_{ab}\varepsilon_{\dot{a}\dot{b}}(\gamma^\mu)_{\alpha\beta}P_\mu,\qquad  \{S^{\alpha}_{a\dot{a}}, S^{\beta}_{b\dot{b}}\} = 2i\varepsilon_{ab}\varepsilon_{\dot{a}\dot{b}} (\gamma^\mu)^{\alpha\beta}K_\mu \ec 
	\cr 
	\{Q_{\alpha a\dot{a}}, S^{\beta}_{b\dot{b}}\} & =  2\left( \varepsilon_{ab}\varepsilon_{\dot{a}\dot{b}}\left(\frac{1}{2}(\gamma^{\mu\nu})_\alpha{}^\beta M_{\mu\nu}  + \delta_{\alpha}^{\beta} D \right) + \delta_{\alpha}^{\beta} \left( \varepsilon_{\dot{a}\dot{b}}R_{ab} - \varepsilon_{ab}\bar{R}_{\dot{a}\dot{b}}\right) \right) \ec 
	\cr
	[R_a{}^b, Q_{\alpha c\dot{c}}] &= \delta_c^b Q_{\alpha a \dot{c}} - \frac{1}{2}\delta_a^b Q_{\alpha c \dot{c}} ,\qquad  [R_a{}^b, S^{\alpha}_{c\dot{c}}] = \delta_c^b S^{\alpha}_{a\dot{c}} - \frac{1}{2}\delta_a^b S^{\alpha}_{c\dot{c}}  \ec 
	\cr
	[\bar{R}_{\;\dot a}^{\dot b}, Q_{\alpha c\dot{c}}] &= -\delta_{\dot c}^{\dot b} Q_{\alpha c \dot{a}} + \frac{1}{2}\delta_{\dot a}^{\dot b} Q_{\alpha c \dot{c}} ,\qquad  [\bar{R}_{\;\dot a}^{\dot b}, S^{\alpha}_{c\dot{c}}] = -\delta_{\dot c}^{\dot b} S^{\alpha}_{c\dot{a}} + \frac{1}{2}\delta_{\dot a}^{\dot b} S^{\alpha}_{c\dot{c}} .
\end{alignat}

The insertion of a 1/2 BPS Wilson line \cite{Cooke:2015ila} breaks the $\mathfrak{osp}(4|4)$ of the bulk theory down to $\mathfrak{su}(1,1|2)$. The $\mathfrak{su}(1,1)$ generators are those of the 1-dimensional conformal group, {\it i.e.} $\{ D,P\equiv P_0,K\equiv K_0\} $, satisfying
\begin{equation}
 [P,K]=-2 D, \qquad [D,P]=P, \qquad [D,K]=-K.
\end{equation}
The preserved R-symmetry is taken to be $\mathfrak{su}(2)_A$ generated by ${R_a}^b$. Since the translations along the line are preserved (as well as special conformal transformations and rotations around the line $M_{12}$), one can conveniently choose $(\gamma^{\mu})_{\alpha}^{\;\;\beta} = (\sigma^{z},\sigma^{x}, \sigma^{y})_{\alpha}^{\;\;\beta}$ as a basis and therefore $(\gamma^{\mu})_{\alpha\beta}=(\sigma^1, -\sigma^3, i\mathbb{1})$ and $(\gamma^{\mu})^{\alpha\beta}=(-\sigma^1, \sigma^3, i\mathbb{1})$. From the anticommutation relations one can identify the preserved fermionic charges. In our conventions
\bea
\{Q_{+ a\dot{a}} , Q_{- b\dot{b}}\} &= 2i\varepsilon_{ab}\varepsilon_{\dot{a}\dot{b}}P.
\eea
 If one picks $Q_{+ a\dot{2}}$, then the closure of the 1-dimensional superconformal algebra requires that the other set of conserved charges be $\epsilon^{ab}Q_{- b\dot{1}}$, and similarly for the superconformal ones. The full set of conserved supercharges is then
\begin{equation}
 Q_a\equiv Q_{+a\dot{2}}, \qquad S_a\equiv iS^{-}_{a\dot{2}},  \qquad  \bar{Q}^a \equiv i\epsilon^{ab}Q_{- b\dot{1}}, \qquad  \bar{S}^a \equiv  -\epsilon^{ab}S^{+}_{b\dot{1}},
\end{equation}
which obey the anticommuation relations
\begin{align}
 \{Q_a , \bar{Q}^b\} &= 2\delta_a^b P ,\qquad  \{S_a, \bar{S}^b\} = 2\delta_a^b K,\cr
 \{Q_{a},\bar{S}^{b}\}&= 2\delta^{b}_{a}\left( D +J_{0} \right) - 2R_{a}^{\;b},\cr
 \{\bar{Q}^{a},S_{b}\}&= 2\delta^{a}_{b}\left(D - J_{0} \right)+ 2R^{\;a}_{b} ,
\end{align}
with 
\bea
 J_{0}= i M_{12}-\bar{R}^{\dot{1}}_{\;\dot{1}}.
\eea

The mixed bosonic/fermionic commutation relations are
\begin{align}
\label{mixedFB1}
[D,{Q_a}]&=\frac{1}{2}{Q_a},\quad  \quad [D,{\bar{Q}^a}]=\frac{1}{2}{\bar{Q}^a}, \quad \quad [D,{S_a}]=-\frac{1}{2}{S_a},\quad  \quad [D,\bar{S}^a]=-\frac{1}{2}\bar{S}^a,\cr
[K,{Q_a}]&={S_a},\quad  \quad [K,{\bar{Q}^a}] ={\bar{S}^a}, \quad \quad [P,{S_a}]=-{Q_a},\quad  \quad [P,\bar{S}^a]=-\bar{Q}^a,
\end{align}
and
\begin{align}
\label{mixedFB2}
[R_a{}^b, Q_{c}] &= \delta_c^b Q_{a} - \frac{1}{2}\delta_a^b Q_{c} ,\qquad  [R_a{}^b, S_{c}] = \delta_c^b S_{a} - \frac{1}{2}\delta_a^b S_{c}  \ec 
	\cr
 [R_a{}^b, \bar{Q}^{c}] &= -\delta_a^c \bar{Q}^{b} + \frac{1}{2}\delta_a^b \bar{Q}^{c} ,\qquad  [R_a{}^b, \bar{S}^{c}] = -\delta_a^c \bar{S}^{b} + \frac{1}{2}\delta_a^b \bar{S}^{c}  \ec 
	\cr
[\bar{R}_{\;\dot 1}^{\dot 1}, Q_{a}] &=\frac{1}{2} Q_a,\qquad  [M^{12}, Q_a] = -i\frac{1}{2} Q_{a}, \qquad [J_{0},Q_{a}]=0 \ec
\cr
[\bar{R}_{\;\dot 1}^{\dot 1}, S_{a}] &=\frac{1}{2} S_a,\qquad  [M^{12}, S_a] = -i\frac{1}{2} S_{a} ,\qquad [J_{0},S_{a}]=0 \ec
\cr
[\bar{R}_{\;\dot 1}^{\dot 1}, \bar{Q}^{a}] &=-\frac{1}{2} \bar{Q}^{a},\qquad  [M^{12}, \bar{Q}^{a}] = i\frac{1}{2} \bar{Q}^{a} ,\qquad [J_{0},\bar{Q}^{a}]=0 \ec
\cr
[\bar{R}_{\;\dot 1}^{\dot 1}, \bar{S}^{a}] &=-\frac{1}{2} \bar{S}^{a},\qquad  [M^{12}, \bar{S}^{a}] = i\frac{1}{2} \bar{S}^{a} ,\qquad [J_{0},\bar{S}^{a}]=0.
\end{align}
Here one can see that the preserved $\mathfrak{u}(1)_{L}\oplus \mathfrak{so}(2)_{\text{rotations}}$ combine in $\mathfrak{u}(1)_{j_0}\oplus \mathfrak{u}(1)_{\text{aut}}$ where $\mathfrak{u}(1)_{j_0}$ is the non-trivial central ideal and $\mathfrak{u}(1)_{\text{aut}}$ is the outer-automorphism generated by $\Tilde{J}= i M_{12}+\bar{R}^{\dot{1}}_{\;\dot{1}}$, giving for example the relations $[\tilde{J}_{\text{aut}},Q_{a}]=+Q_{a}$ and $[\tilde{J}_{\text{aut}},\bar{Q}^{a}]=-\bar{Q}^{a}$. 
Consistently with \cite{Agmon:2020pde}, the maximal subalgebra of $\mathfrak{osp}(4\vert 4)$ is therefore $ \mathfrak{u}(1)_{j_0}\rtimes \mathfrak{psu}(1,1\vert 2)\rtimes \mathfrak{u}(1)_{\text{aut}}$, where $\mathfrak{psu}(1,1\vert 2)\simeq \mathfrak{su}(1,1\vert 2)/\mathfrak{u}(1)_{j_0}$.


\section{Representations of $\mathfrak{su}(1,1|2)$}
\label{app:representations}

We discuss now the representations of  $ \mathfrak{u}(1)_{j_0}\rtimes \mathfrak{psu}(1,1\vert 2)\rtimes \mathfrak{u}(1)_{\text{aut}}$ \cite{Agmon:2020pde}. First let us notice that the defect superconformal algebra has a $ \mathfrak{su}(1,1)\oplus \mathfrak{su}(2)_{R}\oplus\mathfrak{u}(1)_{j_0}\oplus \mathfrak{u}(1)_{\text{aut}}$ subalgebra. We can therefore label the representations in terms of the the Dynkin labels $[\Delta,j_0;j_1]$ where the $\mathfrak{u}(1)_{\text{aut}}$ charge can be neglected as it is unimportant for the following discussion. For the preserved R-symmetry one can define the Cartan generator $H$ and the associated raising and lowering  operators $E^{\pm}$
\begin{equation}
    H = R_{1}^{\;1}-R_{2}^{\;2}=2R_{1}^{\;1},\quad\quad E^{+}=R_{1}^{\;2},\quad\quad E^{-}=R_{2}^{\;1},
\end{equation}
that satisfy 
\begin{equation}
    [H,E^{\pm}] =\pm E^{\pm}, \quad\quad [E^{+},E^{-}]=2H.
\end{equation}
One can therefore characterize the states and the supercharges in terms of the corresponding quantum numbers. In particular
\begin{align}
\label{chargesoftheQs}
    Q_1: \; [\ft{1}{2},0,1],\quad \quad Q_2:\; [\ft{1}{2},0,-1],\quad \quad 
    \bar{Q}^1: \; [\ft{1}{2},0,-1],\quad \quad\bar{Q}^2:\; [\ft{1}{2},0,1].
\end{align}
The highest-weight state $\ket{\Delta,j_0;j_1}$ is defined from the conditions 
\begin{equation}
    S_a\ket{\Delta,j_0;j_1}=0,\quad\quad \Bar{S}^a\ket{\Delta,j_0;j_1}=0,\quad\quad E^{+}\ket{\Delta,j_0;j_1}=0,
\end{equation}
and long multiplets are obtained by acting on it with $Q_a$, $\bar{Q}^a$, $E^-$ and $P$.
The unitarity bound reads
\begin{equation}
    \Delta\geq \abs{j_0}+\frac{1}{2}j_1,
\end{equation}
which is strictly satisfied by the long multiplets. One can also impose the shortening conditions 
\begin{equation}
    \bar{Q}^a\ket{\Delta,j_0;j_1} =0,
\end{equation}
leading to the two following cases
\begin{align}
    1/4\;\text{BPS}:\quad \Delta &= j_0+\frac{1}{2}j_1,\quad\quad L\Bar{A}[j_0]^{(j_1)}_{\Delta},\cr
    1/2\;\text{BPS}:\quad \Delta &= j_0,\quad\quad\quad \quad \; L\Bar{A}[j_0]^{(0)}_{\Delta},
\end{align}
where we adopt (a simplified version of) the notation of \cite{Agmon:2020pde, Cordova:2016emh} with $[j_0]^{(j_1)}_{\Delta}$ indicating the quantum numbers of the superprimary and the capital letters specifying whether the multiplet is long $L$ ($\bar L$) or short at threshold $A$ ($\bar A$) with respect to $Q_a$ ($\bar Q^a$).

The conjugate ones are instead given by the conditions
\begin{equation}
    {Q}_a\ket{\Delta,j_0;j_1} =0,
\end{equation}
which leads to
\begin{align}
    1/4\;\text{BPS}:\quad \Delta &= -j_0+\frac{1}{2}j_1,\quad\quad A\Bar{L}[j_0]^{(j_1)}_{\Delta},\cr 
    1/2\;\text{BPS}:\quad \Delta &= -j_0,\quad\quad\quad \quad \; A\Bar{L}[j_0]^{(0)}_{\Delta}.
\end{align}
One can explicitly recognize the displacement multiplet as  $L\Bar{A}[1]^{(0)}_{1}$. 


\section{Orthogonality conditions}
\label{app:orthogonality}

In this appendix we provide some details on the orthogonality conditions for the blocks $G_{\Delta}(z)$ which we have used to extract the coefficients of the CPW expansions. 

We begin by recalling that the blocks are the eigenfunctions of the differential operator 
\begin{equation}
\label{Doperatorapp}
\cD= a(z) \partial_z^2 + b(z) \partial_z=(1-z) \;z^2\; \partial_z^2+z (2-z)\;\partial_z
\end{equation}
satisfying the eigenvalue equation $\cD G_{\Delta}(z) = \mathfrak{c}_{\Delta} G_{\Delta}(z)$. As noted in the main text, however, the spectrum is degenerate as also the shadow contributions have the same eigenvalue. One can then define the $\omega$-weighted inner product as
\begin{equation}
    \langle G_{\Delta_1} \vert G_{\Delta_2} \rangle \equiv \oint \frac{dz}{2\pi i } \omega(z) G_{\Delta_1}(z) G_{\Delta_2}(z),
\end{equation}
denoting the block contributions as $\ket{G_{\Delta_i} }$ and the shadow ones as  $\bra{G_{\Delta_j} }$. Defining $\tilde{\Delta}$ to be the shadow dimension, such that $\mathfrak{c}_{\Delta} = \mathfrak{c}_{\tilde{\Delta}}$, one has the orthogonality 
\begin{equation}
    \langle G_{\tilde{\Delta}} \vert G_{{\Delta}} \rangle = \oint \frac{dz}{2\pi i } \omega(z) G_{\tilde{\Delta}}(z) G_{\Delta}(z) = \delta_{\Delta, \tilde{\Delta}}.
\end{equation}
Now one must determine the weight $\omega(z)$. This can be done by rewriting \eqref{Doperatorapp} in Sturm-Liouville form
\begin{equation}
    \cD = -\frac{1}{\omega(z)}\frac{\partial}{\partial z} p(z) \frac{\partial}{\partial z},
\end{equation}
from which one gets
\begin{equation}
    a(z) = -\frac{p(z)}{\omega(z)},\qquad b(z) = -\frac{p'(z)}{\omega(z)},\qquad \frac{b(z)-a'(z)}{a(z)}\omega(z)=\omega'(z),
\end{equation}
and finally
\begin{equation}
\label{omegadensity}
    \omega(z)= -\frac{1}{(1-z)^2},
\end{equation}
where the normalization is fixed by requiring orthonormality. 

\bibliographystyle{utphys2}
\bibliography{refs}

\providecommand{\href}[2]{#2}\begingroup\raggedright\begin{thebibliography}{10}\setlength{\parskip}{1pt}\setlength{\itemsep}{0pt
  plus 0.3ex}

\bibitem{Maldacena:1998im}
J.~M. Maldacena, ``{Wilson loops in large N field theories},''
  \href{http://dx.doi.org/10.1103/PhysRevLett.80.4859}{{\em Phys. Rev. Lett.}
  {\bfseries 80} (1998) 4859--4862},
  \href{http://arxiv.org/abs/hep-th/9803002}{{\ttfamily arXiv:hep-th/9803002}}.

\bibitem{Pestun:2016zxk}
V.~Pestun {\em et~al.}, ``{Localization techniques in quantum field
  theories},'' \href{http://dx.doi.org/10.1088/1751-8121/aa63c1}{{\em J. Phys.
  A} {\bfseries 50} no.~44, (2017) 440301},
  \href{http://arxiv.org/abs/1608.02952}{{\ttfamily arXiv:1608.02952
  [hep-th]}}.

\bibitem{Cooke:2017qgm}
M.~Cooke, A.~Dekel, and N.~Drukker, ``{The Wilson loop CFT: Insertion
  dimensions and structure constants from wavy lines},''
  \href{http://dx.doi.org/10.1088/1751-8121/aa7db4}{{\em J. Phys.} {\bfseries
  A50} no.~33, (2017) 335401},
\href{http://arxiv.org/abs/1703.03812}{{\ttfamily arXiv:1703.03812}}.

\bibitem{Giombi:2017cqn}
S.~Giombi, R.~Roiban, and A.~A. Tseytlin, ``{Half-BPS Wilson loop and
  AdS$_2$/CFT$_1$},''
  \href{http://dx.doi.org/10.1016/j.nuclphysb.2017.07.004}{{\em Nucl. Phys. B}
  {\bfseries 922} (2017) 499--527},
  \href{http://arxiv.org/abs/1706.00756}{{\ttfamily arXiv:1706.00756
  [hep-th]}}.

\bibitem{Giombi:2018qox}
S.~Giombi and S.~Komatsu, ``{Exact correlators on the Wilson loop in
  $\mathcal{N}=4$ SYM: Localization, defect CFT, and integrability},''
  \href{http://dx.doi.org/10.1007/JHEP05(2018)109}{{\em JHEP} {\bfseries 05}
  (2018) 109},
\href{http://arxiv.org/abs/1802.05201}{{\ttfamily arXiv:1802.05201}}.

\bibitem{Giombi:2018hsx}
S.~Giombi and S.~Komatsu, ``{More exact results in the Wilson loop defect CFT:
  bulk-defect OPE, nonplanar corrections and quantum spectral curve},''
\href{http://arxiv.org/abs/1811.02369}{{\ttfamily arXiv:1811.02369}}.

\bibitem{Liendo:2018ukf}
P.~Liendo, C.~Meneghelli, and V.~Mitev, ``{Bootstrapping the half-BPS line
  defect},'' \href{http://dx.doi.org/10.1007/JHEP10(2018)077}{{\em JHEP}
  {\bfseries 10} (2018) 077},
\href{http://arxiv.org/abs/1806.01862}{{\ttfamily arXiv:1806.01862}}.

\bibitem{Gimenez-Grau:2019hez}
A.~Gimenez-Grau and P.~Liendo, ``{Bootstrapping line defects in $\mathcal{N}=2$
  theories},'' \href{http://dx.doi.org/10.1007/JHEP03(2020)121}{{\em JHEP}
  {\bfseries 03} (2020) 121}, \href{http://arxiv.org/abs/1907.04345}{{\ttfamily
  arXiv:1907.04345 [hep-th]}}.

\bibitem{Correa:2019rdk}
D.~H. Correa, V.~I. Giraldo-Rivera, and G.~A. Silva, ``{Supersymmetric mixed
  boundary conditions in AdS$_{2}$ and DCFT$_{1}$ marginal deformations},''
  \href{http://dx.doi.org/10.1007/JHEP03(2020)010}{{\em JHEP} {\bfseries 03}
  (2020) 010}, \href{http://arxiv.org/abs/1910.04225}{{\ttfamily
  arXiv:1910.04225 [hep-th]}}.

\bibitem{Beccaria:2019dws}
M.~Beccaria, S.~Giombi, and A.~A. Tseytlin, ``{Correlators on
  non-supersymmetric Wilson line in $ \mathcal{N}=4 $ SYM and
  AdS$_{2}$/CFT$_{1}$},'' \href{http://dx.doi.org/10.1007/JHEP05(2019)122}{{\em
  JHEP} {\bfseries 05} (2019) 122},
  \href{http://arxiv.org/abs/1903.04365}{{\ttfamily arXiv:1903.04365
  [hep-th]}}.

\bibitem{Beccaria:2021rmj}
M.~Beccaria, S.~Giombi, and A.~A. Tseytlin, ``{Higher order RG flow on the
  Wilson line in $ \mathcal{N} $ = 4 SYM},''
  \href{http://dx.doi.org/10.1007/JHEP01(2022)056}{{\em JHEP} {\bfseries 01}
  (2022) 056}, \href{http://arxiv.org/abs/2110.04212}{{\ttfamily
  arXiv:2110.04212 [hep-th]}}.

\bibitem{Ferrero:2021bsb}
P.~Ferrero and C.~Meneghelli, ``{Bootstrapping the half-BPS line defect CFT in
  N=4 supersymmetric Yang-Mills theory at strong coupling},''
  \href{http://dx.doi.org/10.1103/PhysRevD.104.L081703}{{\em Phys. Rev. D}
  {\bfseries 104} no.~8, (2021) L081703},
  \href{http://arxiv.org/abs/2103.10440}{{\ttfamily arXiv:2103.10440
  [hep-th]}}.

\bibitem{Galvagno:2021bbj}
F.~Galvagno and M.~Preti, ``{Wilson loop correlators in $ \mathcal{N} $ = 2
  superconformal quivers},''
  \href{http://dx.doi.org/10.1007/JHEP11(2021)023}{{\em JHEP} {\bfseries 11}
  (2021) 023}, \href{http://arxiv.org/abs/2105.00257}{{\ttfamily
  arXiv:2105.00257 [hep-th]}}.

\bibitem{Barrat:2021yvp}
J.~Barrat, A.~Gimenez-Grau, and P.~Liendo, ``{Bootstrapping holographic defect
  correlators in $ \mathcal{N} $ = 4 super Yang-Mills},''
  \href{http://dx.doi.org/10.1007/JHEP04(2022)093}{{\em JHEP} {\bfseries 04}
  (2022) 093}, \href{http://arxiv.org/abs/2108.13432}{{\ttfamily
  arXiv:2108.13432 [hep-th]}}.

\bibitem{Barrat:2021tpn}
J.~Barrat, P.~Liendo, G.~Peveri, and J.~Plefka, ``{Multipoint correlators on
  the supersymmetric Wilson line defect CFT},''
  \href{http://dx.doi.org/10.1007/JHEP08(2022)067}{{\em JHEP} {\bfseries 08}
  (2022) 067}, \href{http://arxiv.org/abs/2112.10780}{{\ttfamily
  arXiv:2112.10780 [hep-th]}}.

\bibitem{Barrat:2022eim}
J.~Barrat, P.~Liendo, and G.~Peveri, ``{Multipoint correlators on the
  supersymmetric Wilson line defect CFT. Part II. Unprotected operators},''
  \href{http://dx.doi.org/10.1007/JHEP08(2023)198}{{\em JHEP} {\bfseries 08}
  (2023) 198}, \href{http://arxiv.org/abs/2210.14916}{{\ttfamily
  arXiv:2210.14916 [hep-th]}}.

\bibitem{Cavaglia:2022yvv}
A.~Cavagli\`a, N.~Gromov, J.~Julius, and M.~Preti, ``{Integrated correlators
  from integrability: Maldacena-Wilson line in $ \mathcal{N} $ = 4 SYM},''
  \href{http://dx.doi.org/10.1007/JHEP04(2023)026}{{\em JHEP} {\bfseries 04}
  (2023) 026}, \href{http://arxiv.org/abs/2211.03203}{{\ttfamily
  arXiv:2211.03203 [hep-th]}}.

\bibitem{Beccaria:2022bcr}
M.~Beccaria, S.~Giombi, and A.~A. Tseytlin, ``{Wilson loop in general
  representation and RG flow in 1D defect QFT},''
  \href{http://dx.doi.org/10.1088/1751-8121/ac7018}{{\em J. Phys. A} {\bfseries
  55} no.~25, (2022) 255401}, \href{http://arxiv.org/abs/2202.00028}{{\ttfamily
  arXiv:2202.00028 [hep-th]}}.

\bibitem{Cuomo:2021rkm}
G.~Cuomo, Z.~Komargodski, and A.~Raviv-Moshe, ``{Renormalization Group Flows on
  Line Defects},'' \href{http://dx.doi.org/10.1103/PhysRevLett.128.021603}{{\em
  Phys. Rev. Lett.} {\bfseries 128} no.~2, (2022) 021603},
  \href{http://arxiv.org/abs/2108.01117}{{\ttfamily arXiv:2108.01117
  [hep-th]}}.

\bibitem{Aharony:2022ntz}
O.~Aharony, G.~Cuomo, Z.~Komargodski, M.~Mezei, and A.~Raviv-Moshe, ``{Phases
  of Wilson Lines in Conformal Field Theories},''
  \href{http://arxiv.org/abs/2211.11775}{{\ttfamily arXiv:2211.11775
  [hep-th]}}.

\bibitem{Billo:2023ncz}
M.~Billo', F.~Galvagno, M.~Frau, and A.~Lerda, ``{Integrated correlators with a
  Wilson line in $ \mathcal{N} $ = 4 SYM},''
  \href{http://dx.doi.org/10.1007/JHEP12(2023)047}{{\em JHEP} {\bfseries 12}
  (2023) 047}, \href{http://arxiv.org/abs/2308.16575}{{\ttfamily
  arXiv:2308.16575 [hep-th]}}.

\bibitem{Billo:2024kri}
M.~Bill\`o, M.~Frau, F.~Galvagno, and A.~Lerda, ``{A note on integrated
  correlators with a Wilson line in $\mathcal{N}=4$ SYM},''
  \href{http://arxiv.org/abs/2405.10862}{{\ttfamily arXiv:2405.10862
  [hep-th]}}.

\bibitem{Peveri:2023qip}
G.~Peveri, \href{http://dx.doi.org/10.18452/27524}{{\em {Correlators on the
  Wilson Line Defect CFT}}}.
\newblock PhD thesis, Humboldt U., Berlin, 2023.
\newblock \href{http://arxiv.org/abs/2310.17358}{{\ttfamily arXiv:2310.17358
  [hep-th]}}.

\bibitem{Penati:2021tfj}
S.~Penati, ``{Superconformal Line Defects in 3D},''
  \href{http://dx.doi.org/10.3390/universe7090348}{{\em Universe} {\bfseries 7}
  no.~9, (2021) 348}, \href{http://arxiv.org/abs/2108.06483}{{\ttfamily
  arXiv:2108.06483 [hep-th]}}.

\bibitem{Castiglioni:2022yes}
L.~Castiglioni, S.~Penati, M.~Tenser, and D.~Trancanelli, ``{Interpolating
  Wilson loops and enriched RG flows},''
  \href{http://arxiv.org/abs/2211.16501}{{\ttfamily arXiv:2211.16501
  [hep-th]}}.

\bibitem{Castiglioni:2023uus}
L.~Castiglioni, S.~Penati, M.~Tenser, and D.~Trancanelli, ``{Wilson loops and
  defect RG flows in ABJM},'' \href{http://arxiv.org/abs/2305.01647}{{\ttfamily
  arXiv:2305.01647 [hep-th]}}.

\bibitem{Castiglioni:2023tci}
L.~Castiglioni, S.~Penati, M.~Tenser, and D.~Trancanelli, ``{Interpolating
  Bremsstrahlung function in ABJM},''
  \href{http://arxiv.org/abs/2312.13283}{{\ttfamily arXiv:2312.13283
  [hep-th]}}.

\bibitem{Drukker:2009hy}
N.~Drukker and D.~Trancanelli, ``{A supermatrix model for $\cN=6$ super
  Chern-Simons-matter theory},''
  \href{http://dx.doi.org/10.1007/JHEP02(2010)058}{{\em JHEP} {\bfseries 02}
  (2010) 058},
\href{http://arxiv.org/abs/0912.3006}{{\ttfamily arXiv:0912.3006}}.

\bibitem{Aharony:2008ug}
O.~Aharony, O.~Bergman, D.~L. Jafferis, and J.~Maldacena, ``{N=6 superconformal
  Chern-Simons-matter theories, M2-branes and their gravity duals},''
  \href{http://dx.doi.org/10.1088/1126-6708/2008/10/091}{{\em JHEP} {\bfseries
  10} (2008) 091}, \href{http://arxiv.org/abs/0806.1218}{{\ttfamily
  arXiv:0806.1218 [hep-th]}}.

\bibitem{Aharony:2008gk}
O.~Aharony, O.~Bergman, and D.~L. Jafferis, ``{Fractional M2-branes},''
  \href{http://dx.doi.org/10.1088/1126-6708/2008/11/043}{{\em JHEP} {\bfseries
  11} (2008) 043}, \href{http://arxiv.org/abs/0807.4924}{{\ttfamily
  arXiv:0807.4924 [hep-th]}}.

\bibitem{Bianchi:2020hsz}
L.~Bianchi, G.~Bliard, V.~Forini, L.~Griguolo, and D.~Seminara, ``{Analytic
  bootstrap and Witten diagrams for the ABJM Wilson line as defect
  CFT$_{1}$},'' \href{http://dx.doi.org/10.1007/JHEP08(2020)143}{{\em JHEP}
  {\bfseries 08} (2020) 143}, \href{http://arxiv.org/abs/2004.07849}{{\ttfamily
  arXiv:2004.07849 [hep-th]}}.

\bibitem{Ouyang:2015qma}
H.~Ouyang, J.-B. Wu, and J.-j. Zhang, ``{Supersymmetric Wilson loops in $
  \mathcal{N}=4 $ super Chern-Simons-matter theory},''
  \href{http://dx.doi.org/10.1007/JHEP11(2015)213}{{\em JHEP} {\bfseries 11}
  (2015) 213},
\href{http://arxiv.org/abs/1506.06192}{{\ttfamily arXiv:1506.06192}}.

\bibitem{Cooke:2015ila}
M.~Cooke, N.~Drukker, and D.~Trancanelli, ``{A profusion of $1/2$ BPS Wilson
  loops in $\mathcal{N}=4$ Chern-Simons-matter theories},''
  \href{http://dx.doi.org/10.1007/JHEP10(2015)140}{{\em JHEP} {\bfseries 10}
  (2015) 140}, \href{http://arxiv.org/abs/1506.07614}{{\ttfamily
  arXiv:1506.07614 [hep-th]}}.

\bibitem{Ouyang:2015iza}
H.~Ouyang, J.-B. Wu, and J.-j. Zhang, ``{Novel BPS Wilson loops in
  three-dimensional quiver Chern-Simons-matter theories},''
  \href{http://dx.doi.org/10.1016/j.physletb.2015.12.021}{{\em Phys. Lett.}
  {\bfseries B753} (2016) 215--220},
\href{http://arxiv.org/abs/1510.05475}{{\ttfamily arXiv:1510.05475}}.

\bibitem{Ouyang:2015bmy}
H.~Ouyang, J.-B. Wu, and J.-j. Zhang, ``{Construction and classification of
  novel BPS Wilson loops in quiver Chern-Simons-matter theories},''
  \href{http://dx.doi.org/10.1016/j.nuclphysb.2016.07.018}{{\em Nucl. Phys.}
  {\bfseries B910} (2016) 496--527},
\href{http://arxiv.org/abs/1511.02967}{{\ttfamily arXiv:1511.02967}}.

\bibitem{Mauri_2017}
A.~Mauri, S.~Penati, and J.-j. Zhang, ``New bps wilson loops in $ \mathcal{N}=4
  $ circular quiver chern-simons-matter theories,'' {\em Journal of High Energy
  Physics} {\bfseries 2017} no.~11, (Nov, 2017) ,
  \href{http://arxiv.org/abs/1709.03972v3}{{\ttfamily arXiv:1709.03972v3
  [hep-th]}}.

\bibitem{Mauri:2018fsf}
A.~Mauri, H.~Ouyang, S.~Penati, J.-B. Wu, and J.~Zhang, ``{BPS Wilson loops in
  $ \mathcal{N} \geq2$ superconformal Chern-Simons-matter theories},''
  \href{http://dx.doi.org/10.1007/JHEP11(2018)145}{{\em JHEP} {\bfseries 11}
  (2018) 145},
\href{http://arxiv.org/abs/1808.01397}{{\ttfamily arXiv:1808.01397}}.

\bibitem{Gaiotto:2008sd}
D.~Gaiotto and E.~Witten, ``{Janus configurations, Chern-Simons couplings, and
  the $\theta$-angle in ${\cal N}=4$ super Yang-Mills theory},''
  \href{http://dx.doi.org/10.1007/JHEP06(2010)097}{{\em JHEP} {\bfseries 06}
  (2010) 097},
\href{http://arxiv.org/abs/0804.2907}{{\ttfamily arXiv:0804.2907}}.

\bibitem{Imamura:2008dt}
Y.~Imamura and K.~Kimura, ``{N=4 Chern-Simons theories with auxiliary vector
  multiplets},'' \href{http://dx.doi.org/10.1088/1126-6708/2008/10/040}{{\em
  JHEP} {\bfseries 10} (2008) 040},
  \href{http://arxiv.org/abs/0807.2144}{{\ttfamily arXiv:0807.2144 [hep-th]}}.

\bibitem{Hosomichi:2008jd}
K.~Hosomichi, K.-M. Lee, S.~Lee, S.~Lee, and J.~Park, ``{${\cal N}=4$
  superconformal Chern-Simons theories with hyper and twisted hyper
  multiplets},'' \href{http://dx.doi.org/10.1088/1126-6708/2008/07/091}{{\em
  JHEP} {\bfseries 07} (2008) 091},
\href{http://arxiv.org/abs/0805.3662}{{\ttfamily arXiv:0805.3662}}.

\bibitem{Hama:2011ea}
N.~Hama, K.~Hosomichi, and S.~Lee, ``{SUSY gauge theories on squashed
  three-spheres},'' \href{http://dx.doi.org/10.1007/JHEP05(2011)014}{{\em JHEP}
  {\bfseries 05} (2011) 014},
\href{http://arxiv.org/abs/1102.4716}{{\ttfamily arXiv:1102.4716}}.

\bibitem{Drukker:2019bev}
N.~Drukker {\em et~al.}, ``{Roadmap on Wilson loops in 3d Chern-Simons-matter
  theories},'' \href{http://dx.doi.org/10.1088/1751-8121/ab5d50}{{\em J. Phys.
  A} {\bfseries 53} no.~17, (2020) 173001},
  \href{http://arxiv.org/abs/1910.00588}{{\ttfamily arXiv:1910.00588}}.

\bibitem{Agmon:2020pde}
N.~B. Agmon and Y.~Wang, ``{Classifying Superconformal Defects in Diverse
  Dimensions Part I: Superconformal Lines},''
  \href{http://arxiv.org/abs/2009.06650}{{\ttfamily arXiv:2009.06650
  [hep-th]}}.

\bibitem{Semenoff:2004qr}
G.~W. Semenoff and D.~Young, ``{Wavy Wilson line and $AdS$/CFT},''
  \href{http://dx.doi.org/10.1142/S0217751X0502077X}{{\em Int. J. Mod. Phys.}
  {\bfseries A20} (2005) 2833--2846},
\href{http://arxiv.org/abs/hep-th/0405288}{{\ttfamily hep-th/0405288}}.

\bibitem{Drukker:2020dvr}
N.~Drukker, M.~Tenser, and D.~Trancanelli, ``{Notes on hyperloops in $
  \mathcal{N} $ = 4 Chern-Simons-matter theories},''
  \href{http://dx.doi.org/10.1007/JHEP07(2021)159}{{\em JHEP} {\bfseries 07}
  (2021) 159}, \href{http://arxiv.org/abs/2012.07096}{{\ttfamily
  arXiv:2012.07096 [hep-th]}}.

\bibitem{Bianchi:2016vvm}
M.~S. Bianchi, L.~Griguolo, M.~Leoni, A.~Mauri, S.~Penati, and D.~Seminara,
  ``{The quantum 1/2 BPS Wilson loop in ${\cal N}=4$ Chern-Simons-matter
  theories},'' \href{http://dx.doi.org/10.1007/JHEP09(2016)009}{{\em JHEP}
  {\bfseries 09} (2016) 009}, \href{http://arxiv.org/abs/1606.07058}{{\ttfamily
  arXiv:1606.07058}}.

\bibitem{Drukker:2022ywj}
N.~Drukker, Z.~Kong, M.~Probst, M.~Tenser, and D.~Trancanelli, ``{Conformal and
  non-conformal hyperloop deformations of the 1/2 BPS circle},''
  \href{http://dx.doi.org/10.1007/JHEP08(2022)165}{{\em JHEP} {\bfseries 08}
  (2022) 165}, \href{http://arxiv.org/abs/2206.07390}{{\ttfamily
  arXiv:2206.07390 [hep-th]}}.

\bibitem{Liendo:2015cgi}
P.~Liendo, C.~Meneghelli, and V.~Mitev, ``{On Correlation Functions of BPS
  Operators in 3d ${\mathcal{N}}$ = 6 Superconformal Theories},''
  \href{http://dx.doi.org/10.1007/s00220-016-2715-7}{{\em Commun. Math. Phys.}
  {\bfseries 350} no.~1, (2017) 387--419},
  \href{http://arxiv.org/abs/1512.06072}{{\ttfamily arXiv:1512.06072
  [hep-th]}}.

\bibitem{McAvity:1993ue}
D.~M. McAvity and H.~Osborn, ``{Energy momentum tensor in conformal field
  theories near a boundary},''
  \href{http://dx.doi.org/10.1016/0550-3213(93)90005-A}{{\em Nucl. Phys. B}
  {\bfseries 406} (1993) 655--680},
  \href{http://arxiv.org/abs/hep-th/9302068}{{\ttfamily arXiv:hep-th/9302068}}.

\bibitem{Correa:2012at}
D.~Correa, J.~Henn, J.~Maldacena, and A.~Sever, ``{An exact formula for the
  radiation of a moving quark in N=4 super Yang Mills},''
  \href{http://dx.doi.org/10.1007/JHEP06(2012)048}{{\em JHEP} {\bfseries 06}
  (2012) 048}, \href{http://arxiv.org/abs/1202.4455}{{\ttfamily arXiv:1202.4455
  [hep-th]}}.

\bibitem{Billo:2016cpy}
M.~Billo, V.~Goncalves, E.~Lauria, and M.~Meineri, ``{Defects in conformal
  field theory},'' \href{http://dx.doi.org/10.1007/JHEP04(2016)091}{{\em JHEP}
  {\bfseries 04} (2016) 091}, \href{http://arxiv.org/abs/1601.02883}{{\ttfamily
  arXiv:1601.02883 [hep-th]}}.

\bibitem{NagaokaPhD}
G.~Nagaoka,
  \href{http://dx.doi.org/https://doi.org/10.11606/T.43.2022.tde-17082022-132640}{{\em
  {Studies of Wilson Loops on 3d Chern-Simons-Matter Theories}}}.
\newblock PhD thesis, University of S\~ao Paulo, 2022.

\bibitem{Cooke_2019}
M.~Cooke, A.~Dekel, N.~Drukker, D.~Trancanelli, and E.~Vescovi, ``Deformations
  of the circular wilson loop and spectral (in)dependence,''
  \href{http://dx.doi.org/10.1007/jhep01(2019)076}{{\em Journal of High Energy
  Physics} {\bfseries 2019} no.~1, (Jan, 2019) }.
  \url{http://dx.doi.org/10.1007/JHEP01(2019)076}.

\bibitem{Bianchi:2018zpb}
L.~Bianchi, M.~Lemos, and M.~Meineri, ``{Line Defects and Radiation in
  $\mathcal{N}=2$ Conformal Theories},''
  \href{http://dx.doi.org/10.1103/PhysRevLett.121.141601}{{\em Phys. Rev.
  Lett.} {\bfseries 121} no.~14, (2018) 141601},
  \href{http://arxiv.org/abs/1805.04111}{{\ttfamily arXiv:1805.04111
  [hep-th]}}.

\bibitem{Cordova:2016emh}
C.~Cordova, T.~T. Dumitrescu, and K.~Intriligator, ``{Multiplets of
  Superconformal Symmetry in Diverse Dimensions},''
  \href{http://dx.doi.org/10.1007/JHEP03(2019)163}{{\em JHEP} {\bfseries 03}
  (2019) 163}, \href{http://arxiv.org/abs/1612.00809}{{\ttfamily
  arXiv:1612.00809 [hep-th]}}.

\bibitem{Cardinali:2012ru}
V.~Cardinali, L.~Griguolo, G.~Martelloni, and D.~Seminara, ``{New
  supersymmetric Wilson loops in ABJ(M) theories},''
  \href{http://dx.doi.org/10.1016/j.physletb.2012.10.051}{{\em Phys. Lett.}
  {\bfseries B718} (2012) 615--619},
\href{http://arxiv.org/abs/1209.4032}{{\ttfamily arXiv:1209.4032}}.

\bibitem{Bianchi:2014laa}
M.~S. Bianchi, L.~Griguolo, M.~Leoni, S.~Penati, and D.~Seminara, ``{BPS Wilson
  loops and Bremsstrahlung function in ABJ(M): a two loop analysis},''
  \href{http://dx.doi.org/10.1007/JHEP06(2014)123}{{\em JHEP} {\bfseries 06}
  (2014) 123}, \href{http://arxiv.org/abs/1402.4128}{{\ttfamily arXiv:1402.4128
  [hep-th]}}.

\bibitem{Correa:2014aga}
D.~H. Correa, J.~Aguilera-Damia, and G.~A. Silva, ``{Strings in $AdS_4 \times
  \mathbb{CP}^{3}$ Wilson loops in $\mathcal N=$6 super Chern-Simons-matter and
  bremsstrahlung functions},''
  \href{http://dx.doi.org/10.1007/JHEP06(2014)139}{{\em JHEP} {\bfseries 06}
  (2014) 139}, \href{http://arxiv.org/abs/1405.1396}{{\ttfamily arXiv:1405.1396
  [hep-th]}}.

\bibitem{Bianchi_2017}
M.~S. Bianchi, L.~Griguolo, A.~Mauri, S.~Penati, M.~Preti, and D.~Seminara,
  ``Towards the exact bremsstrahlung function of abjm theory,'' {\em Journal of
  High Energy Physics} {\bfseries 2017} no.~8, (Aug, 2017) ,
  \href{http://arxiv.org/abs/1705.10780}{{\ttfamily arXiv:1705.10780
  [hep-th]}}.

\bibitem{Bianchi:2018scb}
L.~Bianchi, M.~Preti, and E.~Vescovi, ``{Exact Bremsstrahlung functions in ABJM
  theory},'' \href{http://dx.doi.org/10.1007/JHEP07(2018)060}{{\em JHEP}
  {\bfseries 07} (2018) 060}, \href{http://arxiv.org/abs/1802.07726}{{\ttfamily
  arXiv:1802.07726 [hep-th]}}.

\bibitem{Dolan:2003hv}
F.~A. Dolan and H.~Osborn, ``{Conformal partial waves and the operator product
  expansion},'' \href{http://dx.doi.org/10.1016/j.nuclphysb.2003.11.016}{{\em
  Nucl. Phys. B} {\bfseries 678} (2004) 491--507},
  \href{http://arxiv.org/abs/hep-th/0309180}{{\ttfamily arXiv:hep-th/0309180}}.

\bibitem{Ferrara:1972uq}
S.~Ferrara, A.~F. Grillo, G.~Parisi, and R.~Gatto, ``{The shadow operator
  formalism for conformal algebra. Vacuum expectation values and operator
  products},'' \href{http://dx.doi.org/10.1007/BF02907130}{{\em Lett. Nuovo
  Cim.} {\bfseries 4S2} (1972) 115--120}.

\bibitem{Poland:2018epd}
D.~Poland, S.~Rychkov, and A.~Vichi, ``{The Conformal Bootstrap: Theory,
  Numerical Techniques, and Applications},''
  \href{http://dx.doi.org/10.1103/RevModPhys.91.015002}{{\em Rev. Mod. Phys.}
  {\bfseries 91} (2019) 015002},
  \href{http://arxiv.org/abs/1805.04405}{{\ttfamily arXiv:1805.04405
  [hep-th]}}.

\bibitem{Polchinski_2011}
J.~Polchinski and J.~Sully, ``Wilson loop renormalization group flows,''
  \href{http://dx.doi.org/10.1007/jhep10(2011)059}{{\em Journal of High Energy
  Physics} {\bfseries 2011} no.~10, (Oct, 2011) }.
  \url{https://doi.org/10.1007%2Fjhep10%282011%29059}.

\bibitem{Drukker:2022pxk}
N.~Drukker, Z.~Kong, and G.~Sakkas, ``{Broken Global Symmetries and Defect
  Conformal Manifolds},''
  \href{http://dx.doi.org/10.1103/PhysRevLett.129.201603}{{\em Phys. Rev.
  Lett.} {\bfseries 129} no.~20, (2022) 201603},
  \href{http://arxiv.org/abs/2203.17157}{{\ttfamily arXiv:2203.17157
  [hep-th]}}.

\bibitem{Bliard:2023zpe}
G.~J.~S. Bliard, \href{http://dx.doi.org/10.18452/27559}{{\em {Perturbative and
  non-perturbative analysis of defect correlators in AdS/CFT}}}.
\newblock PhD thesis, Humboldt U., Berlin, 2023.
\newblock \href{http://arxiv.org/abs/2310.18137}{{\ttfamily arXiv:2310.18137
  [hep-th]}}.

\bibitem{Heemskerk:2009pn}
I.~Heemskerk, J.~Penedones, J.~Polchinski, and J.~Sully, ``{Holography from
  Conformal Field Theory},''
  \href{http://dx.doi.org/10.1088/1126-6708/2009/10/079}{{\em JHEP} {\bfseries
  10} (2009) 079}, \href{http://arxiv.org/abs/0907.0151}{{\ttfamily
  arXiv:0907.0151 [hep-th]}}.

\bibitem{Fitzpatrick:2011dm}
A.~L. Fitzpatrick and J.~Kaplan, ``{Unitarity and the Holographic S-Matrix},''
  \href{http://dx.doi.org/10.1007/JHEP10(2012)032}{{\em JHEP} {\bfseries 10}
  (2012) 032}, \href{http://arxiv.org/abs/1112.4845}{{\ttfamily arXiv:1112.4845
  [hep-th]}}.

\bibitem{Fitzpatrick:2012yx}
A.~L. Fitzpatrick, J.~Kaplan, D.~Poland, and D.~Simmons-Duffin, ``{The Analytic
  Bootstrap and AdS Superhorizon Locality},''
  \href{http://dx.doi.org/10.1007/JHEP12(2013)004}{{\em JHEP} {\bfseries 12}
  (2013) 004}, \href{http://arxiv.org/abs/1212.3616}{{\ttfamily arXiv:1212.3616
  [hep-th]}}.

\bibitem{Gaiotto:2013nva}
D.~Gaiotto, D.~Mazac, and M.~F. Paulos, ``{Bootstrapping the 3d Ising twist
  defect},'' \href{http://dx.doi.org/10.1007/JHEP03(2014)100}{{\em JHEP}
  {\bfseries 03} (2014) 100}, \href{http://arxiv.org/abs/1310.5078}{{\ttfamily
  arXiv:1310.5078 [hep-th]}}.

\bibitem{Ferrero:2023gnu}
P.~Ferrero and C.~Meneghelli, ``{Unmixing the Wilson line defect CFT. Part II:
  analytic bootstrap},'' \href{http://arxiv.org/abs/2312.12551}{{\ttfamily
  arXiv:2312.12551 [hep-th]}}.

\bibitem{Ferrero:2023znz}
P.~Ferrero and C.~Meneghelli, ``{Unmixing the Wilson line defect CFT. Part I.
  Spectrum and kinematics},''
  \href{http://dx.doi.org/10.1007/JHEP05(2024)090}{{\em JHEP} {\bfseries 05}
  (2024) 090}, \href{http://arxiv.org/abs/2312.12550}{{\ttfamily
  arXiv:2312.12550 [hep-th]}}.

\bibitem{Fitzpatrick:2010zm}
A.~L. Fitzpatrick, E.~Katz, D.~Poland, and D.~Simmons-Duffin, ``{Effective
  Conformal Theory and the Flat-Space Limit of AdS},''
  \href{http://dx.doi.org/10.1007/JHEP07(2011)023}{{\em JHEP} {\bfseries 07}
  (2011) 023}, \href{http://arxiv.org/abs/1007.2412}{{\ttfamily arXiv:1007.2412
  [hep-th]}}.

\bibitem{Alday:2014tsa}
L.~F. Alday, A.~Bissi, and T.~Lukowski, ``{Lessons from crossing symmetry at
  large N},'' \href{http://dx.doi.org/10.1007/JHEP06(2015)074}{{\em JHEP}
  {\bfseries 06} (2015) 074}, \href{http://arxiv.org/abs/1410.4717}{{\ttfamily
  arXiv:1410.4717 [hep-th]}}.

\bibitem{Gaiotto:2007qi}
D.~Gaiotto and X.~Yin, ``{Notes on superconformal Chern-Simons-Matter
  theories},'' \href{http://dx.doi.org/10.1088/1126-6708/2007/08/056}{{\em
  JHEP} {\bfseries 08} (2007) 056},
\href{http://arxiv.org/abs/0704.3740}{{\ttfamily arXiv:0704.3740}}.

\bibitem{Lietti:2017gtc}
M.~Lietti, A.~Mauri, S.~Penati, and J.-j. Zhang, ``{String theory duals of
  Wilson loops from Higgsing},''
  \href{http://dx.doi.org/10.1007/JHEP08(2017)030}{{\em JHEP} {\bfseries 08}
  (2017) 030}, \href{http://arxiv.org/abs/1705.02322}{{\ttfamily
  arXiv:1705.02322}}.

\end{thebibliography}\endgroup
\end{document}